\documentclass[pra,aps,english,superscriptaddress]{revtex4-1}
\usepackage[T1]{fontenc}
\usepackage[latin9]{inputenc}
\usepackage{xcolor}
\usepackage{amsmath}
\usepackage{amsthm}
\usepackage{amssymb}
\usepackage{graphicx}
\usepackage{braket}
 \theoremstyle{plain}
 \theoremstyle{definition}
\DeclareMathOperator{\Tr}{Tr}
\DeclareMathOperator{\Real}{Re}
\usepackage[strict]{changepage}

\newtheorem*{stat}{Clausius statement}

\makeatother

\usepackage{babel}

\begin{document}
\title{Quantum thermodynamics of two bosonic systems}
\author{Chiara Macchiavello} 
\affiliation{QUIT Group, Dipartimento di Fisica, Universit\`a di Pavia, via A. Bassi 6, I-27100 Pavia, Italy.}
\affiliation{INFN Sezione di Pavia, via A. Bassi 6, I-27100, Pavia, Italy.}
\affiliation{CNR-INO, largo E. Fermi 6, I-50125, Firenze, Italy.}

\author{Alberto Riccardi}
\affiliation{QUIT Group, Dipartimento di Fisica, Universit\`a  di Pavia, via A. Bassi 6, I-27100 Pavia, Italy.}

\author{Massimiliano F. Sacchi}
\affiliation{CNR - Istituto di Fotonica e Nanotecnologie, 
Piazza Leonardo da Vinci 32, I-20133, Milano, Italy.}
\affiliation{QUIT Group, Dipartimento di Fisica, Universit\`a di Pavia, via A. Bassi 6, I-27100 Pavia, Italy.}

\begin{abstract}
We study the energy exchange between two bosonic systems that interact
via bilinear transformations in the mode operators. The first mode is
considered as the thermodynamic system, while the second is regarded
as the bath. This work finds its roots in a very recent formulation of
quantum thermodynamics \cite{InfoThermo} which allows to consider
baths that are not described by the usual Boltzmann-Gibbs canonical
form. Baths can possess quantum properties, such as squeezing or
coherence, and can be initially correlated with the system, even
through entanglement. We focus mainly on the case of Gaussian states,
by quantifying the relation between their defining parameters, namely
the mean values of the quadratures and the covariance matrix, and
relevant thermodynamical quantities such as the heat exchanged and the
work performed during the interaction process. We fully solve the case
of initially uncorrelated Gaussian states and provide the most general
form of the first law of thermodynamics in this case. We also discuss
the case of initially correlated states by considering a number of
relevant examples, studying how correlations can assist some
phenomena, e.g. work extraction or anomalous heat flows. Finally, we
present an information-theoretic approach based on the Renyi entropy
of order two for clarifying more generally the role of correlations on
heat exchanges.
\end{abstract}
\maketitle

\section{Introduction}
In the last years the theory of thermodynamics, that is the study of
energy and its changes, which are distinguished as heat and work, has
emerged as one of the frameworks where quantum information theory can
be fruitfully employed and, conversely, where new insights on the
theory can be discovered. A new field of research was born, named
quantum thermodynamics \cite{QT1,QT2,QT10,spec}, aiming to extend the
classical thermodynamics to systems of sizes well below the
thermodynamic limit. In this realm, an extensive use of quantum
information tools is applied, ranging from the characterization of heat engines
at the nanoscale \cite{kosl,nano1,nano2} to the study of thermodynamical
properties of relativistic and quantum fields \cite{Fields1,Fields4,Fields2,Fields3}.

\par Standard formulations of thermodynamics,
classical \cite{Callen} or quantum \cite{QT1,QT2,QT10,info1,info2},
consider systems in interaction with thermal baths that are large,
compared to the system dimension, and usually described by states in
the Boltzmann-Gibbs canonical form, whose temperatures are not altered
during thermodynamic processes.  This hypothesis cannot be justified a
priori in the quantum regime, since baths may be small in principle
and can also possess quantum properties, such as coherence
\cite{Coherence1,Coherence2} or squeezing \cite{squeezing1}, and share
correlations
\cite{CorrelationsWinter,Correlations2,Correlations3,Correlations4,Correlations5,Min-energy1,BruschiCor},
e.g.  entanglement, with the system. Energy exchanges between systems and baths will therefore
alter the temperature of both parties and the roles of quantum
properties must be taken into account.  \par To this aim, a
temperature-independent version of thermodynamics has been recently
formulated in \cite{InfoThermo}. Standard thermodynamics can then be
recast as a consequence of information conservation, providing
modifications to the laws of thermodynamics consistent with the
possible presence of (classical or quantum) correlations between
systems and baths, and providing a new definition of the notion of
temperature which generalizes the standard one.  \par According to
such a formulation, in this paper we study the thermodynamics of a
composite system made of two bosonic modes, where one mode represents
the thermodynamical system while the other is treated as the bath,
which can be therefore classically or quantum correlated with the
system and does not have constant and unchangeable temperature. In
particular, we study energy exchanges by considering the first law of
thermodynamics in the case of bilinear interactions of modes.

The paper is organized as follows. In Sec. II we review the approach to
thermodynamics reported in Ref. \cite{InfoThermo}. We focus mainly on
the first law, thus the definitions of internal energy, heat and work
are presented. Then the second law of thermodynamics is briefly
discussed, recalling how the Clausius statement can be generalized in
the presence of initial correlations between the system and the bath.
In Sec. III we discuss how this formulation can be explicitly
expressed for two-mode bosonic states. In particular, we focus on
Gaussian bipartite states and show how the first law of thermodynamics
can be expressed and how the relevant thermodynamical quantities
depend on the state parameters, i.e. the initial mean values of the
quadratures and the covariance matrix, and on the transformations
considered, which are here at most bilinear in the mode operators.  In
Sec. IV we discuss the case when system and bath are initially
uncorrelated; several examples of important classes of two-mode
Gaussian states are presented. In Sec. V we study the case when system
and bath present initial correlations by means of some illustrative
examples, by considering both separable and entangled states. Moreover,
we derive an information-theoretic approach based on the 2-Renyi
entropy in order to clarify the role of correlations on heat
exchanges. This approach allows to link directly the heat flows to the
variations of the amount of correlations.  Finally, in Sec. VI we
consider work extraction schemes, showing how interactions between 
system and bath can be used to increase the free
energy of the system, and thus the amount of extractable work.
\section{Review of quantum thermodynamics with arbitrary bath}
We review here the formulation of thermodynamics introduced in Ref. 
\cite{InfoThermo},
which allows one to go beyond the standard notion of temperature and
to consider baths that possess quantum properties in the most general
form. This section represents therefore a concise synthesis of the
results reported in Ref. \cite{InfoThermo} that we will use in the
next sections of the paper.
\subsection{Intrinsic temperature and complete passivity }
In the standard formulation of thermodynamics one usually considers a
large thermal bath, whose temperature is well defined and whose state
is described by the Boltzmann-Gibbs canonical form. Typically, such a
state is not allowed to change during the thermodynamic processes due
to its ideally infinite thermal capacity.  Hence, for example, after
thermalization the temperature of the system coincides with that of
the thermal bath. In the quantum regime baths can be small in
principle and can also possess quantum properties, such as coherence
and squeezing, or share correlations, e.g. entanglement, with the
system. Therefore, baths cannot be described a priori in the
Boltzmann-Gibbs canonical form and the concept of temperature must be
generalized.  \par This problem has recently been tackled in
\cite{InfoThermo}, where thermodynamics has been viewed as a
consequence of information conservation, leading to the notion of
intrinsic temperature which can be formulated for any quantum system.
The concept of information conservation leads to consider only
operations that are globally entropy-preserving. Hence, given the
state $\rho$ of an arbitrary system, we consider transformations
$\Lambda$ that are entropy-preserving (EP), namely if
$\rho'=\Lambda\left(\rho\right)$ is the state after the action of
$\Lambda$, we have $S\left(\rho\right)=S\left(\rho'\right)$, where
$S\left(X\right)=-\Tr\left[X\log X\right]$ is the von Neumann entropy
of $X$. The central role of entropy can also be seen by the fact that
we establish equivalence classes between states that have the same
entropy, namely we will say that $\rho$ and $\sigma$ belong to the
same class iff $S\left(\rho\right)=S\left(\sigma\right).$ When $H$ is
the Hamiltonian of the system, then we choose as the representative of
each class the state $\gamma\left(\rho\right)$ which has the minimum
energy within the same class of $\rho$, namely
\begin{equation}
\gamma\left(\rho\right)=\arg\min_{\sigma :\ 
  S\left(\sigma\right)=S\left(\rho\right)}E\left(\sigma\right),
\end{equation}
where $E\left(\sigma\right)=\Tr\left[H\sigma\right]$ is the energy of
the state $\sigma.$ \par The min-energy principle
\cite{Min-energy1,Min-energy2,Min-energy3} that allows to find the
state that minimizes the energy at a fixed entropy and is in some
sense complementary to the max-entropy principle
\cite{Max-Entropy1,Max-Entropy2}, indicates that
$\gamma\left(\rho\right)$ must be thermal. Hence, given an entropic
class, the representative thermal state  is given by 
\begin{equation}
\gamma\left(\rho\right)=\frac 1Z e^{-\beta\left(\rho\right)H}
\,,\label{CP state}
\end{equation}
where $Z= \Tr\left[e^{-\beta\left(\rho\right)H}\right]$, and 
$\beta\left(\rho\right)$ represents the intrinsic inverse temperature
of a state $\rho$, which labels the equivalence classes. In this
way we are able to consider a generalized notion of temperature $T(\rho)$ for
any state $\rho,$ not just for the thermal ones, via the relation
\begin{equation}
T\left(\rho\right)=\beta\left(\rho\right)^{-1},    \label{Intrinsic temperature}
\end{equation}
where we fixed the Boltzmann constant as $k_B=1$.  Clearly, for
thermal states one recovers the standard definition of temperature.
The representative state (\ref{CP state}) of each class is also a
completely passive (CP) state \cite{Min-energy2,Min-energy3}, namely
no extractable work can be accessed from it, even when multiple copies
are available. States of this form are also denoted by 
$\gamma\left(H_{S},\beta_{S}\right)$, where $H_{S}$ is the Hamiltonian
of the system and $\beta_{S}$ labels the entropic equivalence
classes, or shortly $\gamma (T_S)$.  
Moreover, these states have also the following property:
given a non-interacting Hamiltonian
$H_{T}=\sum_{X=1}^{N}\mathbb{I}^{\otimes X-1}\otimes
H_{X}\otimes\mathbb{I}^{\otimes N-X}$ , the global $CP$ state is given
by the tensor product of locally CP states with the same inverse
temperature $\beta_{T}$, namely
\begin{equation}
\gamma\left(H_{T},\beta_{T}\right)=\otimes_{X=1}^{N}\gamma\left(H_{X},\beta_{T}\right).
\end{equation}
\subsection{The first law of thermodynamics}
The first law of thermodynamics deals with energy conservation and
describes the energy distribution in terms of variation of heat and
work. We consider a bipartite system described by $\rho_{AB}$, where
$A$ and $B$ represent the thermodynamical system and the bath respectively, that
undergoes an EP transformation that results in the final state $\rho_{AB}'$.
In standard thermodynamics the bath and the system are considered
initially uncorrelated, with the bath in the Boltzmann-Gibbs canonical
form. Heat is then defined as $\Delta
Q=E\left(\rho'_{B}\right)-E\left(\rho_{B}\right)$, corresponding to the
variation of the internal energy of the bath. 
However, this definition of heat
suffers several issues in the quantum domain. For example, if the
temperature of the bath can change and correlations
between system and bath are taken into account, it may lead to seeming violations
of the second law of thermodynamics \cite{Arrow1, Arrow2,Arrow3,Arrow5,CorrelationsWinter}.  Moreover, if we allow the bath to be
non-thermal then its internal energy includes also the share of energy
that can be extracted and converted into work, since only the thermal
states are passive.
\par In the framework of \cite{InfoThermo} two relevant forms of 
energy are distinguished for any given state: the bound and the free
energy. The first represents the amount of internal energy that cannot
be accessed in form of work.  Conversely, the latter is the part of
internal energy that can be transformed into work by an EP
operation. 
\par The bound energy $B\left(\rho\right)$ is defined as
\begin{equation}
B\left(\rho\right)=\min_{\sigma:S\left(\sigma\right)=
  S\left(\rho\right)}E\left(\sigma\right)=E\left(\gamma\left(\rho\right)\right),
\end{equation}
and it represents the energy that cannot be extracted further. The
min-energy principle identifies the energy of the CP state
$\gamma\left(\rho\right)$ as the bound energy. Hence, given a state
$\rho$ and determined its entropic equivalence class, then the energy
of the representative of this class represents the bound energy. 
\par The free energy $F\left(\rho\right)$ is given by the difference
\begin{equation}
F\left(\rho\right)=E\left(\rho\right)-B\left(\rho\right).\label{Free energy}
\end{equation}
This can also be written as
$F\left(\rho\right)=T\left(\rho\right)D\left(\rho||\gamma\left(\rho\right)\right)$,
where
$D\left(\sigma_{1}||\sigma_{2}\right)=\Tr\left[\sigma_{1}\left(\log\sigma_{1}-
  \log\sigma_{2}\right)\right]$ denotes the quantum relative entropy,
which measures the distinguishability of two states $\sigma_{1}$ and
$\sigma_{2}$.  No privileged a priori temperature is required in the
above definition of the free energy. The standard Helmholtz free
energy
$F_{T}\left(\rho\right)=E\left(\rho\right)-TS\left(\rho\right),$ where
$T$ is fixed a priori by choosing a thermal bath in contact with the
system, is recovered in the case of an ideal bath, as was shown in
\cite{InfoThermo}, but Eq.  (\ref{Free energy}) fits in a more general
framework. In standard thermodynamics the Helmholtz free energy allows
to quantify the amount of work $W$ that can be extracted from a system
in contact with an ideal large bath at inverse temperature $\beta$.
Indeed,
$W=F_{\beta}\left(\rho\right)-F_{\beta}\left(\gamma\left(\rho\right)\right),$
where $\gamma\left(\rho\right)$ is the thermal equilibrium state of
the system after the work extraction. The free energy (\ref{Free
  energy}) represents the amount of work that can be extracted by
using a bath at the worst possible temperature, namely
\cite{InfoThermo}
\begin{equation}
F\left(\rho \right)=\min_{\beta} \left[  
F_{\beta}\left(\rho\right)-F_{\beta}\left(\gamma\left(\rho\right) \right)\right].
\label{FFFFF}
\end{equation}
It follows that the inverse temperature that achieves the minimum in
Eq. \eqref{FFFFF} is the intrinsic one $\beta\left(\rho\right)$. 
\par More specifically, given an arbitrary system described by $\rho,$
the extractable work from an EP transformation
$\rho'=\Lambda\left(\rho\right)$, which is
$W=E\left(\rho\right)-E\left(\rho'\right)$, is upper bounded by the
free energy, that is
\begin{equation}
W\leq F\left(\rho\right),\label{wff}
\end{equation}
 where the inequality is saturated iff $\rho'=\gamma\left(\rho\right).$
Moreover, if the global system is in the state $\rho=\rho_{A}\otimes\gamma_{B}\left(T_{B}\right)$
then the right-hand side becomes the standard Helmholtz free energy,
namely in this case $W\leq F_{T_{B}}\left(\rho_{A}\right)-F_{T_{B}}\left(\gamma_{A}\left(T_{B}\right)\right).$ 
\par We can now define heat and work which allow to formulate the first law of
thermodynamics.  Heat represents the most degraded form of energy that
a system exchanges with the bath during thermodynamical processes. It
is therefore natural to define heat as the variation of the bound
energy of the bath, namely \cite{InfoThermo}
\begin{equation}
\Delta Q=B\left(\rho'_{B}\right)-B\left(\rho_{B}\right).\label{Heat dissipated}
\end{equation}
We stress that the bath can present initial correlations with the
thermodynamical system and its intrinsic temperature can change during the
process. If $\Delta Q$ is positive the system dissipates energy during the
process, thus part of its internal energy is transformed into
bound energy of the bath, which contributes to increase its intrinsic
temperature: roughly speaking, we can say that the bath
is heated up during the transformation. Conversely, if $\Delta Q$
is negative the thermodynamical system acquires energy from the bath, whose intrinsic
temperature decreases. 
If the bath is initially in a thermal state $\gamma\left(T_{B}\right)$,
heat is bounded as follows
\begin{equation}
T_{B}\Delta S_{B}\leq\Delta Q\leq\Delta E_{B},
\end{equation}
and the above three quantities coincide in the limit of ideal large thermal
baths. Since the process is entropy-preserving, note that the variation of the von Neumann entropy of the bath
$\Delta S_{B}$ is related to that of the system $\Delta S_{A}$ by the identity
\begin{equation}
\Delta S_{A}+\Delta S_{B}=\Delta I\left(A:B\right),\label{iab}
\end{equation}
 where $\Delta I\left(A:B\right)$ denotes  the variation of the
 mutual information $I\left(A:B\right) = S_A + S_B - S_{AB}$ between
 $A$ and $B$, which measures the amount of both classical and quantum
 correlations shared by the two parties.  
\par The work performed on the thermodynamical 
 system in the EP transformation is $\Delta W_{A}=W-\Delta F_{B}$, 
 where $W$ is the cost needed to implement the EP transformation, namely
 $W=\Delta E_{A}+\Delta E_{B}$, and $\Delta F_B$ represents the
 variation of the free energy of the bath.  
The first law of thermodynamics is then given by 
\begin{equation}
\Delta E_{A}=\Delta W_{A}-\Delta Q.\label{First Law}
\end{equation}
\subsection{ The second law of thermodynamics}
The second law of thermodynamics puts constraints upon the possibility
of some thermodynamical processes providing, for instance, limitations
to the direction of heat exchange or the impossibility of converting
heat into work completely.  For standard thermodynamics it has been
formulated in several equivalent ways, such as the Carnot principle,
the Clausius statement, the Kelvin-Planck statement, or the Caratheodory
principle, just to mention the most known ones. These concepts have
been extended to the more general scenario of entropy-preserving
operations with arbitrary baths in Ref. \cite{InfoThermo}, where the
authors provided generalized statements, which reduce to the usual
ones in the regime of large ideal baths.  Here we recall the
generalized Clausius statement \cite{InfoThermo}:
\begin{stat}
Any iso-entropic process involving two systems $A$
and $B$ in an arbitrary state, with intrinsic temperatures
$T_A$ and $T_B$, respectively, satisfies the following
inequality
\begin{eqnarray}
\left(T_B - T_A \right) \Delta S_A  \geq  \Delta F_A+\Delta F_B  
+ \, T_B \Delta I \left(A : B \right) - W  \label{Second Law},
\end{eqnarray}
where $\Delta F_{X}$ is the change in the free energy of
system $X$, $ \Delta I(A : B)$ is the change of mutual
information, and $W = \Delta E_A + \Delta E_B$ is the
amount of external work performed on the total
system.
\par In the absence of initial correlations between
the two systems, the states being initially
thermal and no external work being performed, this
implies
\begin{eqnarray}
\left( T_B - T_A \right) \Delta S_A \geq  0\,, 
\label{Standard Second Law}
\end{eqnarray}
so that no iso-entropic equilibration process
is possible whose sole result is the transfer of heat
from a cooler to a hotter system.
\end{stat}
\par Equation \eqref{Standard Second Law} may be then overcome for three
reasons: $ (i)$ external work is provided to the global system,
i.e. $W \neq 0 \,$, as in a standard refrigeration cycle; $(ii)$ the
initial states possess free energy that is consumed; $(iii)$ the two
systems are initially correlated. Violations of the standard
formulation due to correlations have recently gained great attention
\cite{Arrow1, Arrow2,Arrow3,Arrow5,CorrelationsWinter}, and several physical systems have been
proposed to test these violations. We will show that the present
framework enables us to study such phenomena for two interacting
Bosonic systems. 
\section{Quantum thermodynamics of two bosonic systems}
\subsection{General description}
We address the study of the first law of thermodynamics (\ref{First
  Law}) for two bosonic systems under  bilinear interaction (for reviews
on the properties of bosonic systems see
\cite{GaussianOlivares,GaussianParis,GaussianAdesso}). Each system is
described by the mode operators $a,a^{\dag}$ and $b,b^{\dag}$,
respectively, with the usual commutation relation, and the total free
Hamiltonian is given by
$H_{0}=H_{A}+H_{B}=\omega_{A}\left(a^{\dag}a+\frac{1}{2}\right)+\omega_{B}\left(b^{\dag}b+\frac{1}{2}\right)$.
The first mode represents the thermodynamical system, while the second
is considered as the bath. We note that two bosonic systems are also
at the basis of analysis of quantum Otto engines \cite{dec} and
thermodynamics of quantum fields \cite{Fields1}.
We will study the following two global transformations
for mode operators
\begin{eqnarray}
&&a'=\cos\theta a+e^{i\varphi}\sin\theta b ,\label{BS mode a-1}\\
&&b'=\cos\theta b-e^{-i\varphi}\sin\theta a ,\label{BS mode b-1}
\end{eqnarray}
with $\theta \in [0,\frac{\pi}{2}]$ and $\varphi \in
[0,2\pi]$, and 
\begin{eqnarray}
&&a'=\cosh
ra+e^{i\psi}\sinh rb^{\dag},\label{S mode a-1}\\
&&b'=\cosh rb + e^{i\psi}\sinh ra^{\dag},\label{S mode b-1}
\end{eqnarray}
with $r\geq0$ and $\psi \in [0,2\pi]$. 
\par The Heisenberg transformation in Eqs. (\ref{BS mode a-1}-\ref{BS mode b-1})
corresponds to a linear mixing of the modes that can describe a frequency
converter for $\omega_{A}\neq\omega_{B}$ or a beam splitter for
$\omega_A =\omega _B $, and is equivalent to the unitary
transformation in the Schroedinger picture \cite{qopt}
\begin{equation}
U_{FC}\left(\zeta \right)=\exp\left\{ \zeta a^{\dag}b-\zeta^{*}ab^{\dag}\right\},
\label{FA}
\end{equation}
with $\zeta = \theta e^{i\varphi }$. 
\par The transformation in Eqs. (\ref{S mode a-1}-\ref{S mode  b-1})
can describe non-degenerate parametric amplification
(i.e. two-mode squeezing), and is equivalent to the unitary
transformation \cite{qopt}
\begin{equation}
U_{PA}\left(\xi \right)=\exp\left\{ \xi a^{\dag}b^{\dag}-\xi^{*}ab\right\} ,
\label{Fcon}
\end{equation}
with $\xi =re^{i\psi}$.
\par The bipartite system undergoes the following process: the initial
state $\rho_{AB}$ is transformed into
$\rho\rq{}_{AB}=U\rho_{AB}U^{\dag}$, where $U=U_{FC}(\zeta)$ or
$U=U_{PA}(\xi )$, and $S\left(\rho'_{AB}\right)=S\left(\rho_{AB}\right)$ being the process
unitary. 
Without loss of generality, the initial state 
can be written as follows
\begin{equation}
\rho_{AB}=D\left(\alpha\right)\otimes
D\left(\delta\right)\xi_{AB}D^{\dag}\left(\alpha\right)\otimes
D^{\dag}\left(\delta\right),\label{stato}
\end{equation}
where $D\left( \lambda \right)= \exp \left( \lambda a^\dag -
\lambda ^* a\right)$ denotes the displacement operator, and $\xi_{AB}$
has zero-mean field values, namely 
$\Tr\left[\left(a\otimes\mathbb{I}_B\right)\xi_{AB}\right]=
\Tr\left[\left(\mathbb{I}_A\otimes b\right)\xi_{AB}\right]=0$.
Notice that the von Neumann entropy of the bath does not depend on the
displacement terms, namely
$S\left(\xi_{B}\right)=S\left(\rho_{B}\right)$.  For the considered
transformations, the final state
$\rho '_{AB}=U\rho_{AB}U^{\dag}$ can also be expressed as
\begin{equation}
\rho '_{AB}=D\left(\alpha ' \right)\otimes
D\left(\delta\rq{}\right)
\xi\rq{}_{AB}D^{\dag}\left(\alpha\rq{}\right)\otimes
D^{\dag}\left(\delta\rq{}\right),\label{StatoFinale}
\end{equation}
where $\xi\rq{}_{AB}= U \xi_{AB} U ^{\dag} $ satisfies
$\Tr\left[\left(a\otimes\mathbb{I}_B\right)\xi\rq{}_{AB}\right]
=\Tr\left[\left(\mathbb{I}_A\otimes
  b\right)\xi\rq{}_{AB}\right]=0$, 
and either 
\begin{eqnarray}
&&\alpha '= \alpha \cos \theta + \delta e^{i\varphi}\sin \theta \,,\\
&&\delta '= \delta \cos\theta -\alpha e^{-i\varphi}\sin\theta \,,
\;
\end{eqnarray}
or 
\begin{eqnarray}
&&\alpha '= \alpha \cosh r + \delta ^*e^{i\psi }\sinh r \,,\\
&&\delta '= \delta \cosh r + \alpha ^*  e^{i\psi}\sinh r \,,
\;
\end{eqnarray}
for transformations $U_{FC}(\zeta)$ and $U_{PA}(\xi)$, respectively.
As a consequence the von Neumann entropy of the bath in the final
state is $S\left(\rho _B ' \right) = S\left(\xi_{B} '
\right)=S\left(\Tr_{A}\left[U\rho_{AB}U^{\dag}\right]\right).$
\par
Our primary aim is now to quantify and discuss the first law $\Delta
W_{A}=\Delta E_{A}+\Delta Q$, with particular focus on heat flows.  In
the present scheme the first law dictates the distribution of energy
between work and heat due to the EP (indeed unitary) interaction.  Let
us first focus on the heat $\Delta
Q=B\left(\rho'_{B}\right)-B\left(\rho_{B}\right)$.  For a thermal
state $\gamma\left(\rho _B\right)$, the von Neumann entropy is given
by $S\left(\gamma\left(\rho_{B}\right)\right)=g\left(N_{B}\right)$,
where $N_{B}=\braket{b^{\dag}b}_{\gamma\left(\rho_{B}\right)}$, and
\begin{eqnarray}
g\left(x\right)=(x+1)\ln\left(x+1\right)-x\ln x\;.\label{gx}
\end{eqnarray}
Hence, being $g(x)$ an
increasing invertible function, one has
$N_{B}=g^{-1}\left[S\left(\gamma\left(\rho_{B}\right)\right)\right]$
and
$E\left(\gamma\left(\rho_{B}\right)\right)=\omega_{B}\left(N_{B}+\frac{1}{2}\right)$,
which represents the bound energy $B(\rho _B)$ for all $\rho_B$ such
that $S(\rho _B)=S(\gamma (\rho _B))$. It follows that $\Delta Q$ can
be expressed as
\begin{equation}
\Delta
Q=\omega_{B}\left[g^{-1}
\left(S\left(\rho'_{B}\right)\right)-g^{-1}\left(S\left(\rho_{B}\right)\right)\right]. \label{Heat
  bosonic}
\end{equation}
\par Let us notice that for infinitesimal transformations Eq. (\ref{Heat
  bosonic}) provides the customary Clausius relation $\delta Q=T_B d
S_B$, where $T_B$ is the intrinsic temperature of the bath. Explicitly,
one has 
\begin{eqnarray}
\delta Q  = \omega _B \frac
       {\partial }{\partial S_B} [g^{-1}(S_B)] dS_B 
=\omega _B \frac {1}{g'(g^{-1}(S_B))}dS_B  = \frac {\omega _B} {
\ln \frac {N_B+1}{N_B}} dS_B = T_B dS_B
\;, \label{inf}
\end{eqnarray}
where we used $g'(x)=\ln \frac{x+1}{x}$ and the identity for the
bosonic Gibbs state $\frac{N_B}{N_B+1}=e^{- \omega /T_B}$.  Moreover,
in the present scenario where total EP transformations are considered,
through Eqs. (\ref{iab}) and (\ref{inf}), the usual separation of the
infinitesimal system entropy variation $dS_A= dS_{rev}+dS_{irr}$
\cite{irr1,irr2,irr3} in terms of exchange (or reversible) entropy
$dS_{rev}=-\frac {\delta Q_B}{T_B}$ and irreversible production of
entropy $dS_{irr}$ allows us to identify the last term as the
variation of mutual information, namely
\begin{eqnarray}
dS_{irr}=dS_A +\frac {\delta Q_B}{T_B} =
dS_A +dS_B= dI(A:B)
\;.
\end{eqnarray}
Hence, the formulation of the second principle of classical
thermodynamics in terms of the statement $dS_{irr}\geq 0$ can be
violated when $dI(A:B) <0$.  \par The variation of the internal energy
in Eq. (\ref{First Law}) is given by
\begin{eqnarray}
\Delta
E_{A}=
\Tr_{AB}\left[\left(H_{A}\otimes\mathbb{I}_{B}\right)\left(\rho'_{AB}-\rho_{AB}\right)\right]
=
\omega_{A}\left 
(\braket{a^{\dag}a}_{\rho_{A}^{'}}-\braket{a^{\dag}a}_{\rho_{A}}\right).\label{Internal
  energy}
\end{eqnarray}
The above relation can be refined by using a phase-space description
of bosonic states. Let us introduce the vector of quadrature operators
$\mathbf{R}=\left(R_{A},R_{B}\right)^{T}=\left(q_{A},p_{A},q_{B},p_{B}\right)^{T},$
where $q_{A}=\frac{1}{\sqrt{2}}\left(a+a^{\dag}\right)$,
$p_{A}=\frac{1}{i\sqrt{2}}\left(a-a^{\dag}\right)$,
$q_{B}=\frac{1}{\sqrt{2}}\left(b+b^{\dag}\right)$, and
$p_{B}=\frac{1}{i\sqrt{2}}\left(b-b^{\dag}\right)$. The components of
$\mathbf{R}$ satisfy
\begin{equation}
\left[R_{k},R_{l}\right]=i\Omega_{kl}, 
\end{equation}
where $\Omega_{kl}$ denotes the element of the symplectic matrix 
\begin{equation}
\Omega = \oplus_{AB} \mathbf{\omega} \,,\label{Omega}
\end{equation}
with 
\begin{equation}
\mathbb{\omega}=\left(\begin{array}{cc}
0 & 1\\
-1 & 0
\end{array}\right).
\end{equation}
For a two-mode bipartite state, the covariance matrix $\sigma_{AB}$,
whose elements are $\sigma_{kl}=\frac{1}{2}\braket{\left\{
  R_{k},R_{l}\right\} }-\braket{R_{k}}\braket{R_{l}}$, can be written
as 
\begin{equation}
\sigma_{AB}=\left(\begin{array}{cc}
\sigma_{A} & \epsilon\\
\epsilon^{T} & \sigma_{B}
\end{array}\right),\label{two-mode Gaussian covariance matrix-1}
\end{equation}
where $\sigma_{A}$ and $\sigma_{B}$ are the covariance 
matrices of the two subsystems and $\epsilon$ describes their correlations. 
The expectation $\braket{\mathbf{R}}_{\rho_{AB}}$ and $\sigma_{AB}$  cannot fully
characterize any two-mode state, since higher-order moments
are generally needed, but they allow to express Eq.  (\ref{Internal
  energy}) as:
\begin{equation}
\Delta
E_{A}=\frac{\omega_{A}}{2}\left(\Tr\left[\sigma'_{A}\right]-\Tr\left[\sigma_{A}\right]+\left\Vert
\braket{R'_{A}}\right\Vert ^{2}-\left\Vert \braket{R_{A}}\right\Vert
^{2}\right),\label{Internal Energy2}
\end{equation}
where $\sigma'_{A}$ and $R'_{A}$ denote the covariance matrix and the
vector of quadrature operators for the thermodynamical system at the
end of the EP transformation, respectively.
\subsection{Two-mode Gaussian states}
States that are fully characterized just by
$\braket{\mathbf{R}}_{\rho_{AB}}$ and $\sigma_{AB}$ are called
Gaussian. These states, named for the Gaussian character of their
Wigner function, can be prepared by applying unitary operators that
are at most bilinear in the mode operators to thermal states. This
feature makes them suitable to be easily prepared and manipulated in
laboratories. Nowadays, several applications are indeed designed
exclusively for Gaussian states in different areas
\cite{GaussianParis,GaussianOlivares,GaussianAdesso,Gaussian3,Gaussian4,Gaussian5}. The
Gaussian character of a state is preserved under linear and bilinear
interaction of the modes. If a two-mode Gaussian state $\rho_{AB}$ is
changed in $\rho'_{AB}=U\rho_{AB}U^{\dag}$ by a unitary $U$ as our
$U_{FC}(\zeta )$ or $U_{PA}(\xi)$ in Eqs. (\ref{FA}) and (\ref{Fcon}),
then there exists a symplectic matrix $\Gamma_{U}$ that transforms
$\mathbf{R}$ and $\sigma_{AB}$ in $\mathbf{R}'=\Gamma_{U} \mathbf{R}$
and $\sigma'_{AB}=\Gamma_{U}\sigma_{AB}\Gamma_{U}^{T}$,
respectively. In this case the symplectic matrix is a $4\times 4$
invertible real matrix satisfying
\begin{equation}
\Gamma\Omega \Gamma^{T}=\Omega,
\end{equation}
where $\Omega$ is defined in (\ref{Omega}). 
The symplectic matrix $\Gamma$ can be decomposed into a block-form as
\begin{equation}
\Gamma=\left(\begin{array}{cc}
A & D\\
C & B
\end{array}\right),\label{symplectic decomposition}
\end{equation}
where $A,B,C$ and $D$ are $2\times 2$ matrices such that:
$A^{T}B-C^{T}D=\mathbb{I}_2$, $A^{T}C=C^{T}A$, and $B^{T}D=D^{T}B$.
\par Equations (\ref{Heat bosonic}) and (\ref{Internal Energy2}) can
be further specified when $\xi_{AB}$ in Eq. (\ref{stato}) is a
Gaussian state and by exploiting the decomposition (\ref{symplectic
  decomposition}).  In fact, any covariance matrix for $n$ modes can
be diagonalized through a symplectic transformation \cite{Wiliamson},
namely $\sigma$ can be written as
\begin{equation}
\sigma=\mathbf{S^{T}WS},\label{symplectic diagonalization-1}
\end{equation}
where $\mathbf{S}$ is a $2n\times2n$ symplectic matrix,
$\mathbf{W}=\oplus_{k=1}^{n}d_{k}\mathbb{I}_{2}$, 
and the elements $d_{k}$ are the symplectic eigenvalues of
$\sigma$.  Heisenberg-Robertson's uncertainty relation imposes
physical constraints on the admissible covariance matrices, which
can be simply expressed in terms of the symplectic eigenvalues as 
\begin{equation}
d_{k}\geq\frac{1}{2},
\label{cons}
\end{equation}
that must hold for any $k$. The symplectic diagonalization (\ref{symplectic diagonalization-1})
implies that any Gaussian state $\rho$ can be obtained from a thermal
state $\nu$ by applying a Gaussian unitary transformation $U_{S}$, associated
to $\mathbf{S}$, namely
\begin{equation}
\rho=U_{S}\nu U_{S}^{\dag},\label{sympl Uni-2}
\end{equation}
where $\nu=\otimes\nu_{k}$ is a product of thermal states $\nu_{k}$. 
\par Let us now determine the intrinsic temperature of a single-mode
Gaussian state. Any such state $\rho$ can be written as 
\begin{equation}
\rho=U_{S}\nu_{N_{th}}U_{S}^{\dag}=D\left(\alpha\right)S\left(\zeta\right)
\nu_{N_{th}}
S\left(\zeta\right)^{\dag}D\left(\alpha\right)^{\dag},\label{Single
  mode gaussian state}
\end{equation}
where $S\left(\zeta\right)=\exp\left [
  \frac{1}{2}\zeta a^{\dag 2}-\frac{1}{2}\zeta^{*}a^{2}\right
]$ denotes the single-mode squeezing operator and $\nu_{N_{th}}$ is a
thermal state with 
$\braket{a^\dag a}_{\nu_{N_{th}}}=N_{th}=(e^{\beta \omega
  _A}-1)^{-1}$.  The unitary operator
$U_{S}$ does not affect the purity $\mu _\rho = \Tr \left[ \rho^{2} \right]
$ and the von Neumann entropy of $\rho$, which therefore depend only
on the thermal seed $\nu _{N_{th}}$ and are given by
\begin{equation}
\mu_{\rho}=\frac{1}{2N_{th}+1},
\end{equation}
and 
\begin{equation}
S\left(\rho\right)=g\left(N_{th}\right),
\end{equation}
respectively. 
Moreover, the purity is related to the determinant of the covariance
matrix of $\rho$ by the relation 
\begin{equation}
\mu_{\rho}=\frac{1}{2\sqrt{\det\left(\sigma_{\rho}\right)}},
\label{purity1}
\end{equation}
and hence
\begin{equation}
N_{th}=\sqrt{\det\left(\sigma_{\rho}\right)}-\frac{1}{2}.\label{Thermal photon}
\end{equation}
As a consequence, the von Neumann entropy depends only on the determinant
of its covariance matrix $\sigma_{\rho}$, and one has
\begin{equation}
S\left(\rho\right)=g\left(\sqrt{\det\left(\sigma_{\rho}\right)} - \frac
12\right).\label{entropy determinant}
\end{equation}
Clearly, the thermal state that represents the entropic equivalence
class of $\rho$ is the one that appears in the decomposition (\ref{Single mode gaussian state}),
which thus identifies the intrinsic temperature of $\rho$ as the
following increasing function of $N_{th}$
\begin{equation}
T\left(\rho\right)=T\left(\nu_{N_{th}}\right)=\omega _A 
\left[\log\left(\frac{1+N_{th}}{N_{th}}\right)\right]^{-1}\,.
\label{tempp}
\end{equation}
\par Notice that for $N_{th}\gg 1$ one has $T\simeq \omega _A N_{th}$. 
Equation (\ref{Thermal photon}) can be used to evaluate the bound
energy of the bath which is in the state $\rho_{B}=\Tr _A [\rho
  _{AB}]$. The thermal state $\gamma\left(\rho_{B}\right)$
representing the entropic class of $\rho_{B}$ has energy 
$\omega_{B}\left(N_{B,th}+\frac{1}{2}\right)$, and hence 
\begin{eqnarray}
  B\left(\rho_{B}\right)=\omega_{B}\left(N_{B,th}+\frac{1}{2}\right)=
  \frac {\omega _{B}}{2} \coth \left ( \frac{\beta \omega _B}{2}
\right) =\omega_{B}\sqrt{\det\left(\sigma_{B}\right)}
\;.
\end{eqnarray}
Notice that for $\beta \omega _{B}\ll 1$ one has $B(\rho _B)\simeq  \beta
^{-1}$, as in the classical equipartition theorem. 
The heat absorbed by the bath is then given by
\begin{equation}
\Delta
Q=B\left(\rho'_{B}\right)-B\left(\rho_{B}\right)=
\omega_{B}\left(\sqrt{\det\left(\sigma'_{B}\right)}-\sqrt{\det\left(\sigma{}_{B}\right)}\right),
\label{Heat Gaussian}
\end{equation}
where $\sigma'_{B}$ is the covariance matrix at the end of the 
transformation, which from Eq. (\ref{symplectic decomposition})
can be expressed as:
\begin{equation}
\sigma'_{B}=B\sigma_{B}B^{T}+C\sigma_{A}C^{T}+C\epsilon B^{T}+
B\epsilon^{T}C^{T}.\label{Final CM bath}
\end{equation}
The direction of the
heat flow is determined just by the sign of
$\Delta_{B}=\det\left(\sigma'_{B}\right)-\det\left(\sigma{}_{B}\right)$. 
We also remind that $\sigma_{B}$ and $\sigma'_{B}$ can be equivalently 
referred to $\left(\rho_{B},\xi_{B}\right)$ and $\left(\rho'_{B},\xi'_{B}\right)$,
respectively, since for the property of Eqs. (\ref{stato}) and
(\ref{StatoFinale}) they do not depend
on the displacement terms.\\ 
Let us now consider the variation of the internal energy of the system
$\Delta E_{A}$. 
By using Eqs. (\ref{Internal Energy2}) and (\ref{symplectic
  decomposition}) one has 
\begin{eqnarray}
  \Delta E_{A}=
\frac{\omega_{A}}{2}\left(\Tr\left[A\sigma_{A}A^{T}+D\sigma_{B}D^{T}+A\epsilon
  D^{T}+D\epsilon^{T}A^{T}\right]\right) +
\frac{\omega_{A}}{2}\left(\left\Vert
\braket{AR_{A}+D R_{B}}\right\Vert ^{2} -\left\Vert
\braket{R_{A}}\right\Vert ^{2}
-\Tr\left[\sigma_{A}\right]\right)\,.\label{Internal energy Gaussian}
\end{eqnarray}
We will show in the following how Eqs. (\ref{Heat Gaussian})
and (\ref{Internal energy Gaussian}) can be explicitly evaluated for
the transformations in Eqs. \eqref{FA} and \eqref{Fcon}.
\subsubsection{\bf{Frequency converter/beam splitter}}
The symplectic matrix $\Gamma_{\zeta}$ corresponding to the
transformation $U_{FC}(\zeta)$ in Eq. (\ref{FA}) is given by
\begin{equation}
\Gamma_{\zeta}=\left(\begin{array}{cc}
\cos\theta\mathbb{I}_{2} & \sin\theta R_{\varphi}\\
-\sin\theta R_{\varphi}^{T} & \cos\theta\mathbb{I}_{2}
\end{array}\right),\label{BS symplectic matrix-2}
\end{equation}
where $R_{\varphi}$ is the rotation operator
\begin{equation}
R_{\varphi}=\left(\begin{array}{cc}
\cos\varphi & \sin\varphi\\
-\sin\varphi & \cos\varphi
\end{array}\right).
\end{equation}
The block-matrix decomposition of $\Gamma_{\zeta}$ is then provided
by: $A=B=\cos\theta\mathbb{I}_{2}$, $D=\sin\theta R_{\varphi}$, and
$C=-\sin\theta R_{\varphi}^{T}$. Hence, the variation of the internal
energy (\ref{Internal energy Gaussian}) for the initial state $\rho_{AB}$
in Eq. (\ref{stato}) reads
\begin{eqnarray}
\Delta E_{A}= 
\omega_{A}\sin^{2}\theta\left[\frac{1}{2}\left(\Tr\left[\sigma_{B}\right]-\Tr\left[\sigma_{A}\right]\right)+\left|\delta\right|^{2}-\left|\alpha\right|^{2}\right]
+ \omega_{A}\sin2\theta\left[\frac{1}{2}
  \Tr\left[R_{\varphi}^T\epsilon\right]+\Real\left(\alpha \delta ^*
  e^{-i\varphi}\right)\right]\,.
\label{internal energy FQ}
\end{eqnarray}
Correspondingly, from Eq. (\ref{Final CM bath}), the covariance matrix of the bath 
evolves as
\begin{equation}
  \sigma _{B}'=\cos^{2}\theta\sigma_{B} + \sin^{2}\theta
  R_{\varphi}^{T}\sigma_{A}R_{\varphi}
  -\frac{1}{2}\sin{2\theta}
  \left(\epsilon^{T}R_{\varphi}+R_{\varphi}^{T}\epsilon\right).\label{BS
    sigma' b}
\end{equation}
Thus, $\Delta Q$ can be computed according to Eq. \eqref{Heat Gaussian}. 
\subsubsection{\bf{Parametric amplifier}}
The symplectic matrix $\Gamma_{\xi}$ corresponding to the
transformation $U_{PA}(\xi)$ in Eq. (\ref{Fcon}) is given by  
\begin{equation}
\Gamma_{\xi}=\left(\begin{array}{cc}
\cosh r\mathbb{I}_{2} & \sinh r\tilde{R}_{\psi}\\
\sinh r\tilde{R}_{\psi} & \cosh r\mathbb{I}_{2}
\end{array}\right),\label{BS symplectic matrix-1-1}
\end{equation}
with 
\begin{eqnarray}
\tilde{R}_{\psi}=\left(\begin{array}{cc}
\cos\psi & \sin\psi\\
\sin\psi & -\cos\psi
\end{array}\right).
\;
\end{eqnarray}
The block-form of $\Gamma_{\xi}$ is expressed by $A=B=\cosh r\mathbb{I}_{2}$
and $C=D=\sinh r\tilde{R}_{\psi}$. Hence, the variation of the internal
energy (\ref{Internal energy Gaussian}) for  the initial state
(\ref{stato}) is
\begin{eqnarray}
  \Delta E_{A}= 
\omega_{A}\sinh^{2}r\left[\frac{1}{2}\left(\Tr\left[\sigma_{B}\right]+\Tr\left[\sigma_{A}\right]\right)+\left|\delta\right|^{2}+\left|\alpha\right|^{2}\right] +   \omega_{A}\sinh2r\left[\frac{1}{2}
  \Tr\left[\tilde{R}_{\psi}\epsilon\right]+\Real\left(\alpha\delta
  e^{-i\psi}\right)\right]\,.  \label{Internal  energy frequency am}
\end{eqnarray}
From Eq. (\ref{Final CM bath}), 
the covariance matrix of the bath after the transformation is 
\begin{eqnarray}
  \sigma'_{B}= \cosh^{2}{r}
\sigma_{B} +
\sinh^{2}r\tilde{R}_{\psi}\sigma_{A}\tilde{R}_{\psi}
+ \frac{1}{2}\sinh{2r}
\left(\epsilon^{T}\tilde{R}_{\psi}+\tilde{R}_{\psi}\epsilon\right)\,.\label{sbp}
\end{eqnarray}
\section{Uncorrelated system and bath}
We consider first the case where the thermodynamical system and the bath are initially
uncorrelated, namely the initial state is
$\rho_{AB}=\rho_{A}\otimes\rho_{B}$, 
being $\rho_{A}$ and $\rho_{B}$ single-mode Gaussian states of the
general form
\begin{eqnarray}
&&\rho_{A}=D\left(\alpha\right)S\left(\zeta_{A}\right)\nu_{N_{A}}S\left(\zeta_{A}\right)^{\dag}
D\left(\alpha\right)^{\dag},\label{StatoAgenerale}
\\ &&
\rho_{B}=D\left(\delta\right)S\left(\zeta_{B}\right)\nu_{N_{B}}S
\left(\zeta_{B}\right)^{\dag}D\left(\delta\right)^{\dag},\label{StatoBgenerale}
\;
\end{eqnarray}
where $\alpha,\delta\in\mathbb{C}$ and
$\zeta_{A}=r_{A}e^{i\theta_{A}}$,  
$\zeta_{B}=r_{B}e^{i\theta_{B}}$,  with $r_{A},r_{B}\geq0$
and $\theta_{A},\theta_{B}\in\left[0,2\pi\right]$. The respective covariance
matrices can be represented in terms of their elements as follows
\cite{GaussianParis}
\begin{eqnarray}
&&\sigma_{X,11}=\frac{2N_{X}+1}{2} \left( \cosh 2r_X + \cos \theta_X
\sinh 2 r_X 
\right), \nonumber \\& &
\sigma_{X,22}=\frac{2N_{X}+1}{2} \left( \cosh 2 r_X - \cos \theta_X \sinh 2r_X  \right),
\nonumber \\& &
\sigma_{X,12}= \sigma_{X,21} = - \frac{2N_{X}+1}{2} \sin \theta_X \sinh 2r_X ,
\;
\end{eqnarray}
where $X=A,B$ labels the mode.  The intrinsic temperatures depend
only on $N_X$ and are obtained by Eq. (\ref{tempp}).
\par Clearly, excluding initial correlations between the modes simplifies
the problem and precludes possible interesting features.  On the
other hand, it allows to analyze and emphasize some quantum properties
of the bath, such as the presence of squeezing that
cannot be found in the standard treatments. For both the considered bilinear
transformations, we will discuss the first law in general.
Particular emphasis will be given to the sign of the heat, i.e. the
direction of heat flow. 
\subsection{Frequency converter/beam splitter}
For the transformation in Eq. \eqref{FA} the variation of the
internal energy (\ref{internal energy FQ}) for a factorized state 
$\rho_{AB}=\rho_{A}\otimes\rho_{B}$, with $\rho_{A}$ and $\rho_{B}$
given by Eqs. (\ref{StatoAgenerale}) and (\ref{StatoBgenerale}), can
be expressed as
\begin{eqnarray}
\Delta E_{A}=
\omega_{A}\sin^{2}\theta\left(\frac{2N_{B}+1}{2}\cosh
2r_{B}-\frac{2N_{A}+1}{2}\cosh 2r_{A}+ \left|\delta\right|^{2}-
\left|\alpha\right|^{2}\right)
+\omega_{A}\sin2\theta\Real\left(\alpha
\delta ^* e^{-i\varphi}\right)
\,,
\label{internal energy FQ-1}
\end{eqnarray}
since the traces of the initial covariance matrices are
$\Tr\left[\sigma_{A}\right]=\left(2N_{A}+1\right)\cosh 2r_{A}$ and
$\Tr\left[\sigma_{B}\right]=\left(2N_{B}+1\right)\cosh 2r_{B}$.  The
internal energy of the system increases the more the bath is squeezed
while, conversely, squeezing in the initial state of the system
decreases the internal energy. The same consideration holds for the
thermal part: the hotter is the bath, namely the higher is its
intrinsic temperature, the more the internal energy increases, while
the opposite holds for the system.  The second term in
Eq. \eqref{internal energy FQ-1} is a phase-sensitive contribution due
to the coherence interference.
\par From Eq. \eqref{BS sigma' b} one has 
\begin{equation}
  \sigma'_{B}=\cos^{2}\theta\sigma_{B} +\sin^{2}\theta \sigma_A^{\varphi},\label{CMB NoCor2}
\end{equation}
 with 
$\sigma_A^{\varphi} = R_{\varphi}^{T}\sigma_{A}R_{\varphi}$. 
The determinant of $\sigma _B '$ can be expressed as \cite{foot}
\begin{eqnarray}
  \det\left(\sigma'_{B}\right)= \cos^4 \theta \det \sigma_B +\sin^4
  \theta \det \sigma_A^{\varphi}
  +\sin^2 \theta \cos^2 \theta
  \left( \sigma^{\varphi}_{A,11}\sigma_{B,22}+
  \sigma^{\varphi}_{A,22}\sigma_{B,11}-2\sigma^{\varphi}_{A,12} \sigma_{B,12}
  \right).
\end{eqnarray}
Explicitly, one obtains 
\begin{eqnarray}
  \det\left(\sigma'_{B}\right)
=  \sin^{4}\theta \left(N_{A}+\frac 12 \right)^2 +
\cos^{4} \theta \left(N_{B}+\frac 12\right)^2
+ 2\sin^{2}\theta \cos^{2} \theta \left(N_{A}+\frac
12\right)\left(N_{B}+\frac 12\right)  F_S,
\label{Det-4} 
\end{eqnarray}
where
\begin{eqnarray}
  F_S
  = \cosh 2r_A \cosh 2r_B  \label{FF}
  -\sinh 2r_{A}\sinh 2r_{B}
  \cos \left( \theta_{AB}-2\varphi \right),
\;
\end{eqnarray}
with $\theta_{AB}=\left(\theta_{A}-\theta_{B}\right)$,
describes how the final temperature of the bath depends on the
squeezing terms. When no squeezing is present $F_S =1$, which is
also its lower bound. Conversely, there is no upper bound. Notice also
that the relative direction of squeezing for the modes 
deeply contributes to the
final temperature, and then to the heat exchanged. 
\par The heat exchanged $\Delta Q$ can be obtained using
Eqs. (\ref{Heat Gaussian}), (\ref{Det-4}), and (\ref{FF}), along with the
relation $\sqrt{\det\left(\sigma{}_{B}\right)}=N_{B}+\frac 12$. 
Since $F_S\geq 1$, then $\sqrt {\det\left(\sigma'_{B}\right)}
\geq   \sin^{2}\theta \left(N_{A}+\frac 12\right) +\cos^{2} 
\theta \left(N_{B}+\frac 12\right)$, and hence 
\begin{eqnarray}
\Delta Q \geq \omega _B \sin ^2 \theta (N_A -N_B)\;. \label{cond}
\end{eqnarray}
In particular, for $\omega _A = \omega _B$ then $W=0$ and no anomalous
heat flows can occur if system and bath are
initially uncorrelated, namely $\Delta Q >0$ iff $T_A >
T_B$. 
\par For simplicity, let us consider $\varphi=0$ and analyze how $F_S$
depends on $\theta_{AB}$, namely on the relative squeezing direction
of the input states. If the modes are squeezed in the same direction,
i.e. $\theta_A =\theta_B$, then $F_S = \cosh \left( 2r_A - 2r_B \right)$
depends just on the difference between the squeezing strengths, and it
may give a small contribution if $r_A \simeq r_B $, even if $r_A,r_B
\gg 1$. If the modes are squeezed in orthogonal directions,
i.e. $\theta_{AB}=\pi $, then $F_S = \cosh \left( 2r_A + 2r_B \right)$,
namely the squeezing strongly increases the final temperature of the
bath, since the two effects add up.  For fixed values of $r_A$ and
$r_B $, the strongest contribution can be achieved when the modes are
squeezed in orthogonal directions.
For arbitrary phase squeezing $\theta _A$ and $\theta _B$ one can
always tune the phase $\varphi$ of the transformation in order to
achieve one of the two above opposite effects.  
\par By combining $\Delta E_{A}$ given in Eq. (\ref{internal energy FQ-1}) and
$\Delta
Q=\omega_{B}\left(\sqrt{\det\left(\sigma'_{B}\right)}-\sqrt{\det\left(\sigma{}_{B}\right)}\right)$,
we can express the work performed on the system as
\begin{equation}
\Delta W_A = \Delta E_A + \Delta Q\,.
\label{MostGeneralWork NoCor}
\end{equation}
Equation \eqref{MostGeneralWork NoCor} provides the most general
formulation of the first law of thermodynamics between two
uncorrelated Gaussian modes which undergo a frequency converter/beam
splitter transformation.
\par We now discuss some illustrative examples to show how Eq.
\eqref{MostGeneralWork NoCor} can be used to study heat flows and the
balance between the different forms of energy.
\subsubsection{Coherent thermal states}
We consider local thermal states with coherent signal $\alpha $ and
$\delta $. The variation of the internal energy of the
system is given by
\begin{eqnarray}
\Delta E_A = \omega_A \sin^2 \theta \left[ \left(N_B -N_A \right) +
  \left| \delta \right|^2 - \left| \alpha \right|^2 \right]
+  \omega_A \sin 2\theta \Real\left(\alpha \delta ^* e^{-i\varphi}\right)\,,
\label{Energy Ex1}
\end{eqnarray}
whereas the heat $\Delta Q$ reads
\begin{equation}
\Delta Q = \omega_B \sin^2\theta \left( N_A -N_B \right).
\label{Heat Ex1}
\end{equation}
The process is then dissipative, i.e. $\Delta Q >0$, iff $N_A > N_B$.
The work performed on
the system is
\begin{eqnarray}
  \Delta W_A = \left( \omega_A - \omega_B \right) \sin^2 \theta
\left(N_B -N_A \right)  
+ \omega_A  \sin^2 \theta \left( \left| \delta
\right|^2 - 
\left| \alpha \right|^2 \right) \nonumber  + \omega_A \sin 2\theta
\Real\left(\alpha\delta ^*
e^{-i\varphi}\right)\,.\label{Work Ex1}
\end{eqnarray}
Note that for a beam splitter, i.e. $\omega = \omega_A=\omega_B$, the work
performed on the system does not depend on the temperature, and one
has
\begin{align}
\Delta W_A =\omega  \sin^2 \theta \left( \left| \delta \right|^2 -
\left| \alpha \right|^2 \right)+ \omega \sin 2\theta
\Real\left(\alpha \delta ^* e^{-i\varphi}\right).
\label{Work Ex1BS}
\end{align}
Moreover, no anomalous heat flows can
occurs, i.e. $\Delta Q >0$ iff $T_A>T_B$. Note also that even if $\left| \alpha \right| =
\left| \delta \right|$ we may have $\Delta W_A\neq0$ for the
interference contribution of the last term in Eq. \eqref{Work Ex1BS}.
\subsubsection{Squeezed states under a balanced frequency converter}
We consider here initial states with $r_A = r_B $, $\alpha=
\delta =0$, and $N_A \neq 0$, $N_B\neq 0$, transformed by balanced frequency
conversion with no phase shift, i.e. $\theta =
\frac{\pi }{4}$ and $\varphi=0 $.  From Eq. (\ref{internal  energy
  FQ}) the variation of the internal
energy of the system reads 
\begin{align}
\Delta E_A =& \frac{\omega_A}{2} \left[ \left(N_B -N_A \right)\cosh 2 r_A \right].
\end{align}
At the end of the transformation, from Eq. (\ref{Det-4}), one has 
\begin{eqnarray}
 \det \sigma'_B = \frac{1}{4}
 \left(N_{A}+ \frac 12\right)^2+\frac{1}{4}
 \left(N_{B}+\frac 12\right)^2  +  \frac{1}{2}
 \left(N_{A}+\frac 12\right) \left(N_{B}+\frac 12 \right) F_S\,,
\end{eqnarray}
where $F_S = \cosh^2 2 r_A - \cos \theta_{AB} \sinh^2 2r_A $. The heat
flows from the bath to the system iff $ \det \sigma'_B - \det \sigma_B
<0$. Since $\det \sigma_B = \left(\frac{2N_{B}+1}{2}\right)^2 $, a
negative heat can be achieved iff 
\begin{equation}
N_B >\frac{1}{6}\left( F_S+2 N_A F_S -3 \right) 
+ \frac{1}{6} \sqrt{(3+F_S^2)(1+2N_A)^2}. \label{71}
\end{equation}
Since $F_S \geq 1$, notice that Eq. (\ref{71}) implies $N_B \geq N_A$, and
hence heat flow from the bath to the
system is not possible if $N_A > N_B$. 
\subsubsection{Squeezed states with phase compensation}
Let us consider the case $N_A \neq 0$, $N_B\neq 0$,
$\theta_{AB}=2\varphi$, and $r_{A}=r_{B}$, namely squeezed thermal
initial states with the relative direction of equal squeezing that matches
the phase of the transformation. The variation of the internal energy of the
system reads as
\begin{eqnarray}
\Delta E_A = \omega_A \sin^2 \theta \left[ \left(N_B -N_A
  \right)\cosh 2r_A + \left| \delta \right|^2 - \left| \alpha \right|^2
  \right]  +  \omega_A \sin 2\theta
\Real\left(\alpha \delta ^*e^{-i\varphi}\right)\,.
\end{eqnarray}
The heat exchanged in the process [see Eq. \eqref{Det-4} 
to compute the determinant in the final state] is
\begin{align}
\Delta Q= \omega_B \sin^{2}\theta\left(N_{A}-N_{B}\right), 
\end{align}
and the inequality (\ref{cond}) is saturated since $F_S =1$.  
Then, the direction of the heat flow is governed just by the
condition $ N_A \lessgtr N_B$, as the states were just thermal. 
This fact can also be
understood from the identity 
\begin{eqnarray}
[U_{FC}(\theta e^{i\varphi}),S(r e^{i\varphi _A})\otimes S(r e^{i\varphi
    _B})]=0\;\label{comm1}
\end{eqnarray}
for $2\varphi =\theta_A -\theta _B$. 
\par The work performed on the system is given by
\begin{eqnarray}
  \Delta W_{A}= \left(\omega_{A}\cosh
  2r_A-\omega_{B}\right)\sin^{2}\theta\left(N_{B}-N_{A}\right) 
  +\omega_{A}\sin^{2}\theta\left(\left|\delta\right|^{2}-\left|\alpha\right|^{2}\right)+
  \omega_{A}\sin2\theta\Real\left(\alpha \delta ^*
  e^{-i\varphi}\right)\,.
\label{Work S1-1}
\end{eqnarray}
Let us analyze the role of the coherent signal $\alpha $ of the
system in the sign of (\ref{Work S1-1}).  For the sake of simplicity
let us put $\delta=0$. Then Eq. (\ref{Work S1-1}) rewrites
\begin{eqnarray}
\Delta W_{A}=  \left(\omega_{A}\cosh 2r_A-\omega_{B}\right)\sin^{2}\theta\left(N_{B}-N_{A}\right)
 - \omega_{A}\sin^{2}\theta\left|\alpha\right|^{2}\,.
\end{eqnarray}
The more the system is displaced, the less work is needed to perform the process. 
 No work is performed on the system for
\begin{equation}
\left|\alpha\right|^{2}=\frac{\left(\omega_{A}\cosh 2r_A-\omega_{B}\right)\left(N_{B}-N_{A}\right)}{\omega_{A}},
\end{equation}
which holds for any $\theta$. Clearly, when $\delta\neq0$ a trade-off
relation between $\alpha$ and $\delta$ emerges, along with
interference effects.
\subsubsection{Equal initial purity}		
We consider initial states with $N_A=N_B$ (i.e. $T_A/T_B= \omega
_A/\omega _B$). The variation of the
internal energy reads 
\begin{eqnarray}
  \Delta E_{A}=  \omega_{A}\sin^{2}\theta \left
         [\left(N_A+ \frac 12\right)
  \left( \cosh 2r_{B}-\cosh 2r_{A}\right) 
+\left|\delta\right|^{2}
-\left|\alpha\right|^{2}\right ] +\omega_{A}\sin2\theta\Real\left(\alpha\delta
^* e^{-i\varphi}\right) \,,\label{internal energy FQ-1TT}
\end{eqnarray}
whereas the determinant of the final covariance matrix of the bath is given by
\begin{equation}
\det\left(\sigma'_{B}\right)=\left(N_{A}+ \frac 12
\right)^{2}\Big[1+ 2\sin^{2}\theta \cos^{2}\theta \left(F_S -1\right) \Big]\,.
\end{equation}
Since $F_S \geq 1$, the process always heats up the bath or at most
$\Delta Q =0$, when $r_A =r_B $ and $\theta_{AB}=2\varphi$. Note that
in this last case the work $\Delta W_A$ will depend only on the
coherent signals. As we will show in Sec. V, the presence of initial
correlations changes drastically the results of this example.  For
$r_A=r_B$ and $2\varphi = \theta _{AB} +\pi$, when $\theta =\pi/ 4$
all the free energy due to the squeezing of both the signal and the
bath is consumed to generate entanglement. This fact can also be
understood by means of the following algebraic identity \cite{tww}
\begin{eqnarray}
 U_{FC}\left( \frac {\pi}{4} e^{i\varphi } \right ) \left ( 
S(r e^{i\theta  _A})\otimes S(r e ^{i\theta _B}) \right )
U^\dag _{FC} \left( \frac \pi 4 e^{i\varphi }\right )
 = U_{PA} 
\left( i r e^{\frac i2 (\theta _A + \theta _B)} \right )
\;,\label{alg}
\end{eqnarray}
for $2\varphi = \theta_A -\theta_B +\pi$.
\subsection{Parametric amplifier}
We now consider the parametric amplifier transformation of
Eq. \eqref{Fcon} for initial product 
state $\rho_{AB}=\rho_{A}\otimes\rho_{B}$, with $\rho_{A}$ and
$\rho_{B}$ given by Eqs. (\ref{StatoAgenerale}) and
(\ref{StatoBgenerale}), respectively. 
The variation of the internal energy of the system
can be written as
\begin{eqnarray}
\Delta
  E_{A}=\omega_{A}\sinh^{2}r\left[ \left
    (\frac{2N_{B}+1}{2}\right )\cosh
    2r_{B}+\left(\frac{2N_{A}+1}{2}\right )\cosh 2r_{A}
    +\left|\delta\right|^{2}+\left|\alpha\right|^{2}
    \right ]
  +\omega_{A}\sinh2r\Real\left(\alpha\delta
  e^{-i\psi}\right)   
  \,.
\label{IEnergy Ncor FA} 
\end{eqnarray}
By increasing the initial squeezing $\Delta  E_{A}$ increases, and the
same occurs by
raising the temperature of both the system and the bath.  
\par From Eq. (\ref{sbp}), the final covariance matrix of the bath is
given by 
\begin{equation}
  \sigma'_{B}=\cosh^{2}r\sigma_{B}+ \sinh^{2}r\tilde{R}_{\psi}\sigma_{A}\tilde{R}_{\psi},
\end{equation}
and its determinant can be expressed as 
\begin{eqnarray}
\det\left(\sigma'_{B}\right)=  \sinh^{4}r \left(\frac{2N_{A}+1}{2}\right)^2+\cosh^{4} r \left(\frac{2N_{B}+1}{2}\right) ^{2}  
+ 2\sinh^{2}r \cosh^{2}r \left(\frac{2N_{A}+1}{2}\right)\left(\frac{2N_{B}+1}{2}\right) G_S\,,
\label{Det-4-1}
\end{eqnarray}
where, similarly to Eq. (\ref{FF}) for $F_S$, one has 
\begin{eqnarray}
G_S= \cosh 2r_A \cosh 2r_B   -
  \sinh 2r_{A}\sinh 2r_{B}
  \cos\left(\theta_{AB}-2\psi\right)\,.
\end{eqnarray}
Since $G_S\geq 1$, then $\sqrt {\det \sigma ' _B}\geq \sinh^2 r
\left(N_{A}+\frac 12\right)+\cosh^{2} r \left(N_{B}+\frac 12 \right)$,
and hence 
\begin{eqnarray}
\Delta Q \geq \omega _B \sinh ^2 r (N_A+N_B+1)\;.\label{cond2}
\end{eqnarray}
Then, for system and bath initially uncorrelated, parametric
amplification always increases the intrinsic temperature of the bath
(and, for symmetry also of the system).  This fact highlights the deep
difference between parametric amplification and frequency conversion.
\par Let us consider explicitly the following example. 
\subsubsection{Squeezed states with phase compensation}
We consider here initial thermal states that are squeezed with equal strength $r_{A}=r_{B}$ and
relative direction matched with the phase of the transformation as
$\theta _{AB}=2\psi $. The heat is then given by
\begin{align}
\Delta Q & =\omega_{B}\sinh^{2}r\left(N_{A}+N_{B}+1\right),
\end{align}
and inequality (\ref{cond2}) is saturated. 
Similarly to the case of frequency conversion, 
when the transformation achieves phase compensation 
the heat is the same as the modes were in thermal
states. 
This fact can also be understood from the
identity
\begin{eqnarray}
[U_{PA} (r e^{i\psi }),S(r'e^{i\theta _A})\otimes S(r'e^{i\theta  _B})]=0\;\label{comm2}
\end{eqnarray}
for $2\psi =\theta_A -\theta _B$, which can be compared with Eq. (\ref{comm1}). \\
The variation of the internal energy reads
\begin{eqnarray}
\Delta E_{A}=\omega_{A}\sinh^{2}r\left(N_{A}+N_{B}+1\right)\cosh
2r_{A}
+\omega_{A}\sinh^{2}r\left(\left|\delta\right|^{2}+\left|\alpha\right|^{2}\right)+\omega_{A}\sinh2r\Real\left(\alpha\delta
e^{-i\psi}\right)\,,\label{IEnergy Ncor  FA-1}
\end{eqnarray}
and hence the work performed on the system is given by 
\begin{eqnarray}
\Delta W_{A}=\sinh^{2}r\left(N_{A}+N_{B}+1\right)\left(\omega_{A}\cosh
2r_{A}+\omega_{B}\right) 
+\omega_{A}\sinh^{2}r\left(\left|\delta\right|^{2}+\left|\alpha\right|^{2}\right)
+\omega_{A}\sinh2r\Real\left(\alpha\delta
e^{-i\psi}\right)\,.\label{WORK  equalSq}
\end{eqnarray}
We notice that for increasing values of the initial squeezing an
increasing work is performed on the system.
\section{Correlated system and bath}
We now examine the case when system and bath are initially
correlated. First, we review the main properties of the correlation
matrices for bipartite Gaussian states. As we will see a complete
description of the correlation matrices is far to be simple and a full
treatment of the thermodynamics in the most general case is beyond the
scope of this paper. Therefore we will consider only some relevant
classes of correlated Gaussian states in order to show how the
presence of correlations affects the results we derived in the
previous Section. In particular, we will see how Eqs.  (\ref{cond})
and (\ref{cond2}) can be violated.  Finally, the problem of heat
exchanges in the presence of correlations is discussed from an
information-theoretic perspective by using the Renyi entropy of order
$2$.
\subsection{Correlations for bipartite Gaussian states}
The covariance matrix of two bosonic modes is generally given as in
Eq. \eqref{two-mode Gaussian covariance matrix-1}. The constraints in
Eq. \eqref{cons} can be written as
\begin{equation}
d_{\pm}\geq\frac{1}{2}\,,\label{cdpm}
\end{equation}
where $d_{\pm}$ are the symplectic eigenvalues that can be computed
by 
\begin{equation}
  d_{\pm}^{2}=
  \frac{\Delta\left(\sigma_{AB}\right)\pm\sqrt{\Delta\left(\sigma_{AB}\right)^{2}-4\det\sigma_{AB}}}{2},
\end{equation}
with
$\Delta\left(\sigma_{AB}\right)=\det\sigma_{A}+\det\sigma_{B}+2\det\epsilon$.
Here $\det\sigma_{AB}$ and $\Delta\left(\sigma_{AB}\right)$ are global
symplectic invariants, while $\det\sigma_{A}$, $\det\sigma_{B}$, and
$\det\epsilon$ are local symplectic invariants. \\ The matrix
$\epsilon$ encodes the information about the correlations, which can also
reveal the presence of entanglement. In the case of a two-mode
Gaussian state, the positive partial transpose (PPT) criterion
\cite{PeresPPT}, named in this case Simon\rq{}s criterion, provides a
necessary and sufficient condition for entanglement
\cite{SimonPPT,PirandolaCorrelationMatrix}.  Indeed, given a two-mode
Gaussian state $\rho_{AB}$ with covariance matrix $\sigma_{AB}$, the
state is entangled iff $\sigma_{AB}$ is positive definite,
$d_{-}\geq\frac{1}{2}$ and $\tilde{d}_{-}<\frac{1}{2}$, where
$\tilde{d}_{-}$ is the symplectic eigenvalue of the partial transposed
covariance matrix $\tilde{\sigma}_{AB}$, which is given by 
\begin{equation}
\tilde{d}_{-}=\sqrt{\frac{\tilde{\Delta}-\sqrt{\tilde{\Delta}^{2}-4\det\sigma_{AB}}}{2}},
\end{equation}
with $\tilde{\Delta}=\det\sigma_{A}+\det\sigma_{B}-2\det\epsilon$.
Furthermore, a necessary, but not sufficient, condition for
entanglement is $\det\epsilon<0$. 
\par Generally, when
one is only interested in the correlations properties of a 
bipartite system, a different way of writing the covariance
matrix $\sigma_{AB}$ is helpful. In fact, any two-mode covariance matrix can
be brought into a normal form via local symplectic transformations,
namely for any $\sigma_{AB}$ there exists a symplectic matrix
$\mathbf{S_{N}}=\mathbf{S_{A}\oplus S_{B}}$, with $\mathbf{S_{A}}$ and
$\mathbf{S_{B}}$ acting on the first and second
mode respectively, such that the transformed matrix
$\sigma_{AB}^{N}=\mathbf{S_{N}^{T}}\sigma_{AB}\mathbf{S_{N}}$ can be
expressed as
\begin{equation}
\sigma_{AB}^{N}=\left(\begin{array}{cccc}
a & 0 & c_{+} & 0\\
0 & a & 0 & c_{-}\\
c_{+} & 0 & b & 0\\
0 & c_{-} & 0 & b
\end{array}\right),
\end{equation}
where $\det\sigma_{A}=a^{2}$, $\det\sigma_{B}=b^{2}$,
$\det\epsilon=c_{+}c_{-}$, and
$\det\sigma_{AB}=\left(ab-c_{+}^{2}\right)\left(ab-c_{-}^{2}\right)$.
Such four real parameters, which are uniquely determined, up to a
common sign flip between $c_{-}$ and $c_{+}$, allow to study
the correlations between the parties in an easy and correct way, since local
transformations do not change the amount of correlations. From a
thermodynamical study, however, this approach is not justified, since
local squeezing and their relative direction play 
an important role in energy exchanges, as we saw in the
previous section. 
\par On the other hand, the treatment of two-mode Gaussian states in the
most general form involves too many parameters, thus a full study of
the thermodynamics in such case is beyond the scope of this paper. Our
aim is to show that in the presence of correlations new phenomena
arise and to set up the framework for future work. Then, in the
following we will limit our study only to local thermal states,
correlated in two possible ways, characterized by the choices:
$c_{+}=c_{-}=c$ (Type-I) and $c_{+}=-c_{-}=c$ (Type-II),
respectively. Type-I class contains only separable states, while
Type-II can also describe entangled states.  This approach will give
us general hints about the role of correlations in the studied 
thermodynamical processes.
\subsection{Type-I correlated states}
The covariance matrix for Type-I correlated states is given by
\begin{equation}
\sigma_{AB}^{I}=\left(\begin{array}{cccc}
\frac{2N_{A}+1}{2} & 0 & c & 0\\
0 & \frac{2N_{A}+1}{2} & 0 & c\\
c & 0 & \frac{2N_{B}+1}{2} & 0\\
0 & c & 0 & \frac{2N_{B}+1}{2}
\end{array}\right).\label{CM CLASS}
\end{equation}
The constraints of Eq. (\ref{cdpm}) impose that $c$ is bounded as
\begin{equation}
\left|c\right|\leq\sqrt{N_{A}N_{B}}.\label{System2}
\end{equation}
The coherent contribution to the modes is described by the complex parameters 
$\alpha$ and $\delta$  for system and bath, respectively. All states
belonging to this class are not entangled.
\subsubsection{Frequency converter/beam splitter}
The variation of the internal energy of the thermodynamical system for this class of
states can be computed according to Eq. (\ref{internal energy
  FQ}). Since $\epsilon=c\mathbb{I}_{2}$, one has
\begin{eqnarray}
\Delta E_{A}= 
\omega_{A}\sin^{2}\theta\left(N_{B}-N_{A}+\left|\delta\right|^{2}-\left|\alpha\right|^{2}\right)
+ \omega_{A}\sin2\theta\left[c\cos\varphi+\Real\left(\alpha\delta
  ^* e^{-i\varphi}\right)\right]\,.\label{CC internal energy phi}
\end{eqnarray}
From Eq. (\ref{BS sigma' b}) one also has 
\begin{eqnarray}
  \sigma'_{B}=\Big(\frac{2N_{A}+1}{2}\sin^{2}\theta+\frac{2N_{B}+1}{2}\cos^{2}\theta
  -c\sin2\theta\cos\varphi \Big )\mathbb{I}_{2}\,,
\end{eqnarray}
and hence the heat is given by 
\begin{equation}
\Delta
Q=\omega_{B}\left[\left(N_{A}-N_{B}\right)\sin^{2}\theta-c\sin2\theta\cos\varphi\right]\,.
\label{CC
  heat phi}
\end{equation}
Notice that inequality (\ref{cond}) for uncorrelated input states can
now be violated.  
\par The work performed on the system writes
\begin{eqnarray}
\Delta W_{A}=
  \left(\omega_{A}-\omega_{B}\right)\left[\left(N_{B}-N_{A}\right)\sin^{2}\theta+c\sin2\theta\cos\varphi\right]
    +\omega_{A}\left[\sin^{2}\theta\left(\left|\delta\right|^{2}-\left|\alpha\right|^{2}\right)+\sin2\theta\Real\left(\alpha\delta
    ^* e^{-i\varphi}\right)\right]\,. \label{CC work-1}
\end{eqnarray}
Note first that for $\omega =\omega_{A}=\omega_{B}$, i.e. for a beam splitter,
the work performed on the system is independent of the correlations
and one recovers Eq.  \eqref{Work Ex1BS}. Indeed, the increase
(decrease) in the internal energy due to correlations is exactly
balanced by the heat released (absorbed) by the bath. The work in this
case depends only on the coherence terms described by the displacement
operators.\\ Let us assume for simplicity that the coherent signal is
set to zero and $\omega_A = \omega _B$, which implies that $\Delta W_A
= 0 $. Consider now the case when $N_A > N_B$, namely system is hotter
than the bath, and focus on the heat flow. 
As long as condition (\ref{System2}) is also satisfied, the heat is
negative if
\begin{equation}
c>\frac{1}{2}\left(N_{A}-N_{B}\right)\frac{\tan\theta}{\cos\varphi},
\end{equation}
when $\cos \varphi > 0$ (note that $\tan \theta > 0$, since 
$\theta \in [0,\frac{\pi}{2}]$). If $\cos \varphi < 0$,  we have a
  negative heat flow for 
\begin{equation}
c<\frac{1}{2}\left(N_{A}-N_{B}\right)\frac{\tan\theta}{\cos\varphi}.
\end{equation}
This can lead to an apparent violation of the second law:
after the process the bath is colder, even if initially $T_A>T_B$ and
no work is performed on the system. Finally, notice that for $\varphi
= \frac{\pi}{2}$ or $\frac{3\pi}{2}$ the presence of initial
correlations does not affect any thermodynamical quantity. 
\subsubsection{Parametric amplifier}
For initial state with covariance matrix
of the form (\ref{CM CLASS}),  the variation of the internal
energy from Eq. (\ref{Internal  energy frequency am}) is given by 
\begin{eqnarray}
\Delta E_{A}= 
\omega_{A}\sinh^{2}r\left(N_{A}+N_{B}+1+\left|\delta\right|^{2}+\left|\alpha\right|^{2}\right)
+ \omega_{A}\sinh2r \Real\left(\alpha\delta
  e^{-i\psi}\right)\,,\label{Internal energy frequency am-1}
\end{eqnarray}
which does not depend on the correlations,  since $\epsilon = c \,
\mathbb{I}_2$ and so $\Tr [{\tilde R}_{\psi}\epsilon ]=0$. 
\par The final covariance matrix of the bath reads
\begin{eqnarray}
\sigma '_B=\left (\frac{2N_{A}+1}{2}\sinh^{2}r
+\frac{2N_{B}+1}{2}\cosh^{2}r \right ) \mathbb{I}_2 + c\sinh2r \tilde{R}_\psi\,, 
\end{eqnarray}
and hence 
\begin{eqnarray}
\det\sigma_{B}'= 
\left(\frac{2N_{A}+1}{2}\sinh^{2}r+\frac{2N_{B}+1}{2}\cosh^{2}r\right )^{2}
-  c^2\sinh^{2}2r\,.
\end{eqnarray}
The heat exchanged can then be expressed as
\begin{eqnarray}
\Delta Q=
\omega_{B}\sqrt{\left(\frac{2N_{A}+1}{2}\sinh^{2}r+\frac{2N_{B}+1}{2}\cosh^{2}r\right
  )^{2}-c^2\sinh^{2}2r}
-\omega_{B}\frac{2N_{B}+1}{2} \,.\label{QC heat-1} 
\end{eqnarray}
For fixed $N_{A}$ and $N_{B}$ the heat has its maximum value for
$c=0$, for which $\Delta Q = \omega_B \sinh^{2}r \left( N_A + N_B +
1\right) \geq 0$. Notice that such value also saturates inequality
(\ref{cond2}), which holds only for factorized initial states.  For
increasing values of the correlation $ \left| c \right|$ the heat
decreases, and may even become negative, differently from 
the case of Sec. IV B for uncorrelated input states.  In fact, for 
\begin{equation}
\left|c\right|>\frac{\sqrt{\left(\frac{2N_{A}+1}{2}\sinh^{2}r+
\frac{2N_{B}+1}{2}\cosh^{2}r\right)^{2}-\left(\frac{2N_{B}+1}{2}\right)^{2}}}{\sinh
  2r}\,,\label{System1}
\end{equation}
one has $\Delta Q <0$. Note, however, that $\left|c\right|$ cannot be arbitrarily large since
necessarily 
$\left|c\right|\leq\sqrt{N_{A}N_{B}}\equiv c_{M}$ in order to
guarantee a physical state. For instance, if $r=1$, $N_{A}=20$ and
$N_B = 10$, then we have $\Delta Q<0$ if
$13.90<\left|c\right|\leq c_{M}=14.14$. Since
the right-hand side in \eqref{System1} increases with $r$ while
$c$ is bounded, notice also that there exists a maximum value of $r$ for which the
condition \eqref{System1} can be satisfied while keeping 
$N_A$ and $N_B$ fixed. Anyway, the
minimum of $\Delta Q$ versus the correlations is reached for $ |c|=
c_{M}$. 
\par Finally, the work performed on the system is
\begin{eqnarray}
\Delta W_{A}&&= \omega_{A}\sinh2r \Real\left(\alpha\delta
    e^{-i\psi}\right)
  + \omega_{A}\sinh^{2}r\left
  (N_{A}+N_{B}+1+\left|\delta\right|^{2}+\left|\alpha\right|^{2}\right)
  \nonumber \\ 
 && +\omega_{B}\sqrt{\left(\frac{2N_{A}+1}{2}\sinh^{2}r+\frac{2N_{B}+1}{2}\cosh^{2}r\right)^{2}-\left(c\sinh2r\right)^{2}}  -\omega_{B}\frac{2N_{B}+1}{2}\,.
\end{eqnarray}
Note that the dependency of the work on the correlations cannot be
eliminated even for $\omega_{A}=\omega_{B}$, differently from
the case of frequency conversion.  Moreover, for increasing values of
$|c|$, $\Delta W_{A}$ decreases. Therefore, we have shown that the
presence of correlations, although pertaining to separable states,
allows new phenomena for this process, such as negative heat flows,
which are impossible in the absence of correlations. This may happen
even when the bath is colder than the system.
\subsection{Type-II correlated states}
The covariance matrix for Type-II correlated states is given by 
\begin{equation}
\sigma_{AB}^{II}=\left(\begin{array}{cccc}
\frac{2N_{A}+1}{2} & 0 & c & 0\\
0 & \frac{2N_{A}+1}{2} & 0 & -c\\
c & 0 & \frac{2N_{B}+1}{2} & 0\\
0 & -c & 0 & \frac{2N_{B}+1}{2}
\end{array}\right),\label{CM classical/quantum}
\end{equation}
These states are locally thermal and can be separable or
entangled, depending on the range of $c$. The constraints in
Eq. (\ref{cdpm}) impose the conditions
\begin{equation}
\left|c\right|\leq\sqrt{N_{A}\left(1+N_{B}\right)}\,,\label{nab}
\end{equation}
if $N_{A}\leq N_{B}$, or 
\begin{equation}
\left|c\right|\leq\sqrt{N_{B}\left(1+N_{A}\right)}\,\label{nba}
\end{equation}
if $N_{A}>N_{B}$. By applying Simon's criterion, we know that
system and bath are entangled iff 
\begin{equation}
\left|c\right|>\sqrt{N_{A}N_{B}}.\label{Entanglement Bound}
\end{equation}
\subsubsection{Frequency converter/beam splitter}
For this class of states $\Delta E_{A}$, given by Eq. (\ref{internal energy FQ}),
reads
\begin{eqnarray}
\Delta E_{A}=
\omega_{A}\sin^{2}\theta\left(N_{B}-N_{A}+\left|\delta\right|^{2}-\left|\alpha\right|^{2}\right)
+ \omega_A \sin2\theta \Real\left(\alpha
\delta ^* e^{-i\varphi}\right) \,,\label{QC  internal energy}
\end{eqnarray}
which is independent  of the correlations, differently from the
  Type-I states.    In fact, since
$\epsilon=c\sigma_{Z}$, then $\Tr\left[R_{\varphi}\epsilon\right]=0$ 
for any $\varphi$. The final covariance matrix of the bath is
\begin{eqnarray}
\sigma_{B}'=\left ( \frac{2N_{A}+1}{2}\sin^2\theta
+\frac{2N_{B}+1}{2}\cos^{2}\theta \right ) \mathbb{I}_2 -
c\sin2\theta \, {\bar R}_\varphi \,,
\end{eqnarray}
where
\begin{equation}
{\bar R}_\varphi = \left( \begin{array}{cc}
\cos\varphi & -\sin \varphi \\
-\sin \varphi & -\cos \varphi
\end{array} \right),
\end{equation}
and its determinant is given by
\begin{eqnarray}
  \det\sigma_{B}' 
  =\left(\frac{2N_{A}+1}{2}\sin^{2}\theta+\frac{2N_{B}+1}{2}\cos^{2}\theta\right)^{2}-c^2
\sin^{2}2\theta
\;.
\end{eqnarray}
Hence, the stronger are the correlations, the lower is the final temperature of the
bath. Since $\det\sigma_{B}=\left(\frac{2N_{B}+1}{2}\right)^{2}$, 
the heat exchanged can be expressed
as
\begin{eqnarray}
\Delta Q= 
\omega_{B}\sqrt{\left(N_{A}\sin^{2}\theta+N_{B}\cos^{2}\theta+\frac{1}{2}\right)^{2}-\left(c\sin2\theta\right)^{2}}-
\omega_{B}\frac{2N_{B}+1}{2}\,.\label{QC heat}
\end{eqnarray}
The heat $\Delta Q$ has a maximum for $c=0$, for which
inequality (\ref{cond2}) is saturated, and decreases for increasing values of 
$\left|c\right|$. Since for fixed
$N_{A}$ and $N_{B}$ the state is entangled only if $\left|c\right| >
\sqrt{N_A N_B} $, the decrease of $\Delta Q$ is emphasized the more the
initial state is entangled. 
For 
\begin{equation}
\left|c\right|>\frac{\sqrt{\left(N_{A}\sin^{2}\theta+N_{B}\cos^{2}\theta+\frac{1}{2}\right)^{2}
    -\left(\frac{2N_{B}+1}{2}\right)^{2}}}{\sin2\theta}\,,
\end{equation}
along with condition (\ref{nab}) or (\ref{nba}), the heat flow becomes
negative. Anomalous heat flows can be found also in this case, with
general enhancement for increasing correlations/entanglement.  
\subsubsection{Parametric amplifier}
The variation of the internal energy (\ref{Internal  energy frequency
  am})  for initial states with covariance
matrix \eqref{CM classical/quantum} reads
\begin{eqnarray}
\Delta E_{A}= 
\omega_{A}\sinh^{2}r\left(N_{A}+N_{B}+1+\left|\delta\right|^{2}+\left|\alpha\right|^{2}\right)
+  \omega_{A}\sinh2r\left[c\cos\psi+\Real\left(\alpha\delta
  e^{-i\psi}\right)\right]\,.\label{Internal energy frequency am-1-1}
\end{eqnarray}
Since the final covariance matrix of the bath is
$\sigma_{B}'=
\left(\frac{2N_{A}+1}{2}\sinh^{2}r+\frac{2N_{B}+1}{2}\cosh^{2}r+
c\sinh2r\cos\psi\right)\mathbb{I}_{2}$,  the corresponding heat is given by
\begin{equation}
\Delta Q=\omega_{B}\left[\left(N_{A}+N_{B}+1\right)\sinh^{2}r+c\sinh2r\cos\psi\right]\,.
\label{qbb}
\end{equation}
Hence, the work performed
on the system writes 
\begin{eqnarray}  
  \Delta W_{A}&&=
  \omega_{A}\left[\sinh^{2}r\left(\left|\delta\right|^{2}+\left|\alpha\right|^{2}\right)+\sinh2r\Real\left(\alpha\delta
    e^{-i\psi}\right)\right] \nonumber \\& & +
  \left(\omega_{A}+\omega_{B}\right)[
    \left(N_{B}+N_{A}+1\right)\sinh^{2}r + c \sinh2r\cos\psi]\,.
\end{eqnarray}
Both $\Delta E_A$ and $\Delta Q$ depend in
the same way on the correlations, and their effect adds up in the
work. The strength of 
correlations rule the sign of all these thermodynamic quantities.
For instance, we have $\Delta Q <0$ iff
\begin{equation}
c<-\left(\frac{N_{A}+N_{B}+1}{2}\right)\frac{\tanh r}{\cos\psi}\,,
\end{equation}
for $\cos \psi >0$, or 
\begin{equation}
c>-\left(\frac{N_{A}+N_{B}+1}{2}\right)\frac{\tanh r}{\cos\psi}\,,
\end{equation}
for $\cos \psi <0$, with the additional constraint (\ref{nab}) or
(\ref{nba}). 
Stronger correlations, in the sense of greater
values of $\left|c\right|$, do not result automatically in lower
values for $\Delta Q$ and $\Delta W_{A}$, due to a not trivial
dependence on the phase $\psi $. Finally, notice that for $\psi
= \frac{\pi}{2}$ or $\frac{3\pi}{2}$ the presence of initial
correlations does not affect any thermodynamical quantity. 
\subsection{The role of correlations on heat exchanges: an information-theoretic
approach} In the previous subsection we have focused on the role of
initial correlations between system and bath, considering the heat
flow and the work performed on the system. Here, following an
information-theoretic approach, we consider a different perspective by
studying how the variation of correlations between the initial and
final state can determine the heat flows for a general
entropy-preserving transformation. \\ For generic interaction
between two bosonic systems the heat is evaluated by Eq. \eqref{Heat
  bosonic}.  As a consequence, the sign of the heat is determined by
the sign of the variation of the bath von Neumann entropy, namely we
have $\Delta Q >0$ iff $\Delta S_B =
S\left(\rho\rq{}_B\right)-S\left(\rho_B\right) >0$.  Since for
entropy-preserving transformations we have $\Delta I \left( A:B
\right) = \Delta S_A + \Delta S_B$, the heat exchange is positive iff
$\Delta I \left( A:B \right) > \Delta S_A $. When both $\Delta S_A >0
$ and $\Delta S_B >0 $, both the intrinsic temperatures of system and
bath increase, thus leading to an increase in the correlations.  Conversely,
if the process lowers both temperatures, it must also decrease the
total amount of correlations.
\par For bipartite pure
states the von Neumann entropy of the marginal state(s) is a well
defined measure of entanglement. As a consequence, for a pure initial
state one has $\Delta Q > 0$ iff the amount of entanglement increases,
while a decreasing entanglement implies that the temperature of the
bath decreases, independently of the temperature of the system. Notice that this may also happen for 
$W=0$, i.e. when no external work is performed.  
A paradigmatic example is the case of two pure squeezed
states in orthogonal direction under a balanced beam splitter,
producing a twin-beam state, and the reversed transformation. 
\par Specifically for Gaussian states, a quantifier of the information
encoded in a state $\rho$ is provided by the Renyi-2 entropy \cite{Reny2A,Reny2B,Landi1}
\begin{equation}
S_{2}\left(\rho\right)=\frac{1}{2}\ln\left(\det\sigma\right),
\end{equation}
where $\sigma$ is the covariance matrix associated to $\rho$. By
comparing with Eq. \eqref{entropy determinant}, we notice that the
Renyi-2 entropy just replaces the concave function $g(x)$ of
Eq. (\ref{gx}) appearing in the von Neumann entropy by a different
concave function, namely $\ln (x+\frac 12) $. Recalling
Eq. \eqref{purity1}, one also has $S_2 (\rho) = - \ln (2\mu_\rho)$.
\par The Rényi-2 entropy satisfies the strong subadditivity inequality
for all Gaussian states and coincides up to a constant with the Wigner
entropy. Typically, this link allows for a fundamental simplification
of the problem of characterizing entropy production, as one can map an
open system dynamics into a Fokker-Planck equation for the Wigner
function, and hence employ tools of classical stochastic processes to
obtain simple expressions for entropy production rate and entropy flux
rate from the system to the environment
\cite{Landi1,Landi3,Landi4,Landi5,Landi2}.  The Renyi-2 entropy can
also be used to define a Gaussian entanglement measure, the Gaussian
Renyi-2 (GR2) entanglement, which for pure states $\psi_{AB}$ is given
by
\begin{equation}
\mathcal{E}_{2}\left(\psi_{AB}\right)=\frac{1}{2}\ln\left(\det \sigma_{B}\right)=S_{2}\left(\rho_{B}\right).
\end{equation}
The bound energy can be reformulated in terms of the Renyi-2
entropy, namely
$B\left(\chi_{B}\right)=\omega_{B}\exp\left(S_{2}\left(\chi_{B}\right)\right)$,
and hence the heat rewrites 
\begin{equation}
\Delta Q=\omega_{B}\left\{ \exp\left[S_{2}\left(\rho'_{B}\right)\right]-\exp\left[S_{2}\left(\rho_{B}\right)\right]\right\} .
\end{equation}
Then, for pure Gaussian states the heat has a clear interpretation in terms of the variation of the GR2
entanglement, i.e. 
\begin{equation}
\Delta Q=\omega_{B}\left\{
\exp\left[\mathcal{E}_{2}\left(\rho'_{AB}\right)\right]-\exp\left[\mathcal{E}_{2}\left(\rho_{AB}\right)\right]\right\}.
\end{equation}
Here again we see that if the state loses entanglement in the
thermodynamical transformation then $\Delta Q<0$, namely the bath
becomes colder. Conversely, if the
transformation increases the entanglement, the bound energy of the
bath increases. 
\par In the
case of mixed states we expect
that also correlations of separable states are involved in the heat exchange,
as we have already seen in many previous examples. The Renyi-2 entropy 
leads to a well-defined measure of correlations for Gaussian states,
namely the
Renyi-$2$ mutual information
\begin{equation}
I_{2}\left(A:B\right)=\frac{1}{2}\ln\left(\frac{\det\sigma_{A}\det\sigma_{B}}{\det\sigma_{AB}}\right).\label{Mutual info-1}
\end{equation}
Let us consider its variation under Gaussian transformations
\begin{equation}
\Delta
I_{2}\left(A:B\right)=\frac{1}{2}\ln\left(\frac{\det\sigma'_{A}\det\sigma
  '_{B}}{\det\sigma_{A}\det\sigma_{B}}\right),\label{Variation M info}
\end{equation}
where we used the fact that $\det\sigma_{AB}'=\det\sigma_{AB}$. 
Equivalently, one has
\begin{equation}
\Delta I_{2}\left(A:B\right)=\Delta S_{2}\left(\rho_{A}\right)
+\ln\left(\frac{B\left(\rho'_{B}\right)}{B\left(\rho_{B}\right)}\right),
\end{equation}
which implies 
\begin{equation}
B\left(\rho'_{B}\right)= \left[ \exp\Big( 
\Delta
  I_{2}\left(A:B\right)-\Delta S_{2}\left(\rho_{A}\right)\Big)
\right] B\left(\rho_{B}\right)\,.\label{prod}
\end{equation}
The relevant point of equality (\ref{prod}) is the fact that it
relates initial and final temperature of the bath by a multiplicative
factor directly related to information quantities. In this context
then the bath is heated iff $\Delta I_2 \left( A:B \right) > \Delta
S_2 \left(\rho_A \right)$.  Finally, we recall that the Renyi-2 entropy
and mutual information have already been explicitly related to the
entropy production in irreversible processes
\cite{Landi1,Landi3,Landi4,Landi5}, also assessed experimentally
\cite{Landi2}. Hence, further developments may consider irreversible
entropy production mechanisms in the present framework.
\section{Work extraction schemes}
Our final aim is to suggest possible implementations for work
extraction. For a bipartite system, described by a global state
$\rho_{AB}$, the extractable work is upper bounded by the global free
energy $F \left( \rho_{AB} \right)$. In most cases it is not possible
to have full access to both the system and the bath, since one may
perfectly control the former but not the latter. Hence, extracting the
whole $F \left( \rho_{AB} \right)$ might be an impossible task. On the
other hand, if one considers only the system then it is possible to
extract at most $F \left( \rho_A \right)$, which is generally far
below $F \left( \rho_{AB} \right)$. To enhance the amount of
extractable work one can make the system interact with the bath via an
entropy-preserving process, increasing the free energy to $F \left(
\rho_A ' \right) > F \left( \rho_A \right)$, and then extract work
from the system. Here we do not formalize a working medium to
explicitly implement the work extraction from the system, but we will
instead focus on engineering the interaction via bilinear
transformation of the modes pertaining to system and bath, in order to
achieve the optimal increase of the free energy of the system.  \par
In this context, for a fair treatment the relevant quantity is the
variation $\tilde W \equiv \Delta F_A -W $, which represents the
balance between the increase of the free energy of the system and the
cost employed by the interaction $W = \Delta E_A + \Delta E_B$ (which
is zero for the passive beam-splitter). Recalling that the free energy
bounds the extractable work, then a positive value of $\tilde W$ is
intended as a positive contribution for work extraction in the
first-stage of a work engine, where just system and bath interact. We
also notice the possibility of having both $\Delta F_A > 0$ and $W <
0$, which means that an increase of the system free energy can be
contextually accompanied by work extraction \cite{notaw}.  Notice also
the following equivalent expression for $\tilde W$
\begin{eqnarray}
\tilde W = -(\Delta E_B + \Delta B_A)\;,
\label{NetWork}
\end{eqnarray}
which formally corresponds to the work extracted from the bath.  For
the study of $\tilde W$, we will consider bipartite states in
Eq. (\ref{stato}) with no coherent signal on the system, i.e. $\alpha
=0$, since the corresponding contribution to the free energy of the
states can be trivially extracted by the inverse unitary displacement
with no entropy exchange. Since the bath is considered as not directly
accessible, its coherent signal cannot be extracted with the same
procedure, and it must be generally taken into account in the
engineering of the transformation.  \par The net increase in the
extractable work $\tilde{W}$ is given by the two contributions $\Delta
E_B$ and $\Delta B_A$, which for a general two-mode bosonic state can
be expressed as
\begin{eqnarray}
&&\Delta
E_{B}=\omega_{B}\left(\braket{b^{\dag}b}_{\rho_{B}^{'}}-\braket{b^{\dag}b}_{\rho_{B}}\right), \label{Internal
  energyB}
  \\
  &&\Delta B_A=\omega_{A}\left[g^{-1}\left(S\left(\rho'_{A}\right)\right)-g^{-1}\left(S\left(\rho_{A}\right)\right)\right]\,,\label{BoundA}
\end{eqnarray}
respectively. Therefore, to get a positive $\tilde{W}$ a competition appears between
an appropriate reduction of the internal energy of the bath and a sufficient decrease
in the bound energy---and hence entropy---of the system.
\par For Gaussian states undergoing a symplectic
transformation $\Gamma $ as in Eq. \eqref{symplectic decomposition}, the
variation of the internal energy of the bath is given by
\begin{eqnarray}
\Delta E_{B}=
\frac{\omega_{B}}{2}\Tr\left[B\sigma_{B}B^{T}+C\sigma_{A}C^{T}
  +C\epsilon B^{T}+B\epsilon^{T}C^{T} \right] +
\frac{\omega_{B}}{2}\left(\left\Vert \braket{B R_{B}}\right\Vert ^2
-\left\Vert \braket{R_{B}}\right\Vert ^2 -\Tr\left[ \sigma_B
  \right]\right) \,,
\label{Delta EB}
\end{eqnarray} 
where we considered that the system does not have coherent
signal.\\ The variation of the system bound energy is given by $\Delta
B_A = \omega_A \left( \sqrt{\det \sigma\rq{}_A} -\sqrt{\det \sigma_A}
\right)$, where the covariance matrix of the system after the process
reads as
\begin{equation}
  \sigma'_{A}=A\sigma_{A}A^{T} + D\sigma_{B}D^{T}+ 
  +A\epsilon D^{T}+D\epsilon^{T}A^{T}. \label{Delta BA}
\end{equation}
The net increase in the free energy of the system $\tilde{W}$ depends
on the symplectic transformation $\Gamma$ that makes the system
interact with the bath.  Typically, $\Gamma$ is characterized by some
parameters that reflect possible configurations of an experimental
setup. Since $\tilde{W}$ is a function of the transformation,
i.e. $\tilde{W}=\tilde{W} (\Gamma)$,  we can look for the appropriate
configuration of $\Gamma$ such that $\tilde{W}(\Gamma) >0$ and is
maximum. 
\subsection{Frequency converter/beam splitter}
For this process the variation of
the internal energy of the bath is given by 
\begin{eqnarray}
\Delta E_{B}=  \frac{\omega_B}{2}\sin^{2}\theta\left(
\Tr\left[\sigma_{A}\right]-\Tr\left[\sigma_{B}\right] -2\lvert \delta
\rvert ^2  \right) -
 \frac{\omega_B}{2}\sin2\theta\Tr\left[R_{\varphi}^T\epsilon\right]\,, 
\end{eqnarray}
whereas the final covariance matrix of the system reads  
\begin{equation}
\sigma'_{A}=
\cos^{2}\theta\sigma_{A}+\sin^{2}\theta
R_{\varphi}\sigma_{B}R_{\varphi}^T
+\frac{1}{2}\sin{2\theta}
\left(\epsilon R_{\varphi}^T +R_{\varphi}\epsilon ^T\right).
\end{equation}
 Since for this process $\Delta E_B = - \frac{\omega_B}{\omega_A}
 \Delta E_A$ \cite{notaw}, one has $\tilde{W} =
 \frac{\omega_B}{\omega_A} \Delta E_A - \Delta B_A$.  Hence, the
 employed transformation must increase the internal energy of the
 system while decreasing its intrinsic temperature. 
\par In the following we consider specific examples for the frequency
converter transformations.  
\subsubsection{Coherence from the bath} 
In this simple example we consider a state of the form 
\begin{eqnarray}
I\otimes D(\delta ) (\nu _{N_A}\otimes \nu _{N_B}) I\otimes D^\dag
(\delta ), \;\label{dd}
\end{eqnarray}
namely two local thermal modes, with coherent signal in the bath
mode. Using a frequency converter one has 
\begin{equation}
\Delta F_A = \omega_A \sin^2 \theta \lvert \delta \rvert ^2, 
\end{equation}
which has the maximum value $\Delta {F}_{\max}= \omega _A |\delta |^2$
for $U_{FC}\left ( \frac \pi 2 \right )$.
This corresponds to a swap of the system and the bath state, which is
physically not trivial for $\omega _A \neq \omega _B$.
The net increase in the extractable work
$\tilde{W}$ is given by 
\begin{equation}
\tilde{W} = \left( \omega_B - \omega_A \right)\sin^2 \theta \left( N_B
- N_A \right) + \omega_B \sin^2 \theta \lvert \delta \rvert ^2,
\end{equation}
which is positive as long as
\begin{eqnarray}
|\delta |^2 >  \frac{\omega_A - \omega_B }{\omega_B}\left( N_B - N_A \right)
\;,\label{poss}
\end{eqnarray}
and maximum for $\theta =\pi /2$.  If condition (\ref{poss}) is
violated the cost of the transformation exceeds the increase in the
extractable work. We observe that for a balanced beam splitter
(i.e. $\omega_A=\omega _B$ and $\theta =\pi/4$) the transformation
would leave system and bath at the same temperature, but would be less
efficient since it would create correlations and also leave coherence
(and hence free energy) in the bath. Finally, notice that when Eq.
(\ref{poss}) holds along with
\begin{eqnarray}
W=(\omega _A - \omega _B)(|\delta |^2 +N_B -N_A) <0
  \;
\end{eqnarray}
the increase of the system free energy is accompanied by work
extraction. 
\subsubsection{Exploiting type-I correlated states} 
In the following example the optimal transformation strongly depends on the
state parameters. We consider correlated local thermal states with 
$\epsilon = c\, \mathbb{I}_2$, which are therefore
described by the covariance matrix given in \eqref{CM CLASS}. 
We have seen that the
net increase in the free energy of the system is $\tilde{W} =
\frac{\omega_B}{\omega_A} \Delta E_A - \Delta B_A$. According to
Eq. \eqref{CC internal energy phi}, the variation of
the internal energy of the system for this class of states is given by
\begin{eqnarray}
\Delta E_A = \omega_A \sin^{2}\theta \left (N_B - N_A + |\delta |^2
\right) +
\omega_A \sin 2\theta \, c\, \cos \varphi \,,
\;
\end{eqnarray}
whereas the corresponding variation of the bound energy reads 
\begin{equation}
\Delta B_A = \omega_A \sin^{2}\theta \left( N_B -N_A \right) +
\omega_A \sin 2\theta \,c\,\cos \varphi \,.
\end{equation}
Hence,  one obtains
\begin{eqnarray}
\tilde{W} = \left( \omega_B - \omega_A \right) \big[ \sin^{2}\theta
  \left( N_B -N_A \right) + \sin 2\theta \, c\, \cos \varphi  \Big]+
\omega_B \sin^{2}\theta |\delta |^2
\label{WorkEx1}
\;.
\end{eqnarray}
Let us consider the case $\delta =0$ in order to study the pure effect
of correlations.  
For this class of states if $\omega _A =\omega_B$ one has
$\tilde{W}=0$, namely the procedure does not provide any advantage.
\par Suppose now that $\omega_B >
\omega_A$. From Eq. \eqref{WorkEx1} it is clear that the optimal choice of
$\varphi$ depends on the sign of
$c$, namely if $c<0$ then $\varphi=\pi$, while for $c>0$ we choose
$\varphi=0$.  For $N_B>N_A$ we have $\tilde{W}>0$ for any $\theta $. 
If $N_B<N_A$, a positive $\tilde{W}$
can be obtained if
\begin{equation}
\tan \theta < \frac{2 |c|}{N_A-N_B}.
\label{theta1}
\end{equation}
In both cases the optimal value of $\theta$ maximizing $\tilde{W}$ is given by
\begin{equation}
\theta_{\max} = \frac{1}{2} \arctan \left(  \frac{2|c|}{N_A-N_B} \right),\label{theta3}
\end{equation}
with the suitable choice $\varphi =0$ or $\varphi =\pi$. Notice that the
more correlated is the state, the larger is the net increase of the
system free energy.  
\par
\begin{figure}[htb]
{\includegraphics[scale=0.46]{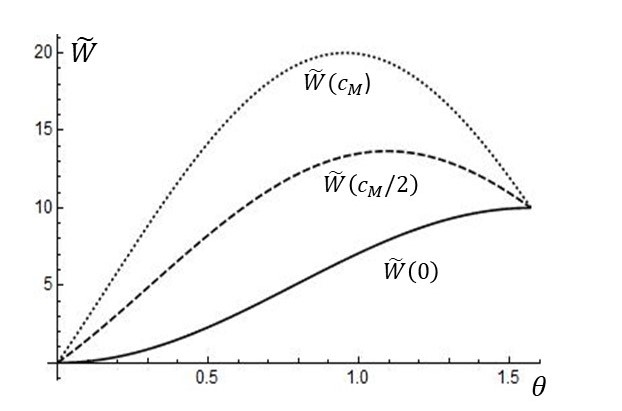}}
\caption{\label{f:fig4} Plots of $\tilde{W}$ for
  classically correlated thermal states with no coherent signal,  $N_A = 5$,
  $N_B=10$, and $\omega_B=2\omega_A$,  for different values of
  $c$. $\tilde{W}\left(c_M \right)$ represents $\tilde{W}$ for the
  maximum value of the correlations $c_M = \sqrt{N_A N_B}$, while
  $\tilde{W}\left(c_M / 2 \right)$ for $c=c_M /2$. $\tilde{W}\left(0
  \right)$ is the extractable work for  uncorrelated modes, which
  is maximized for $\theta =\pi /2 $.}
\end{figure}

\begin{figure}
{\includegraphics[scale=0.4]{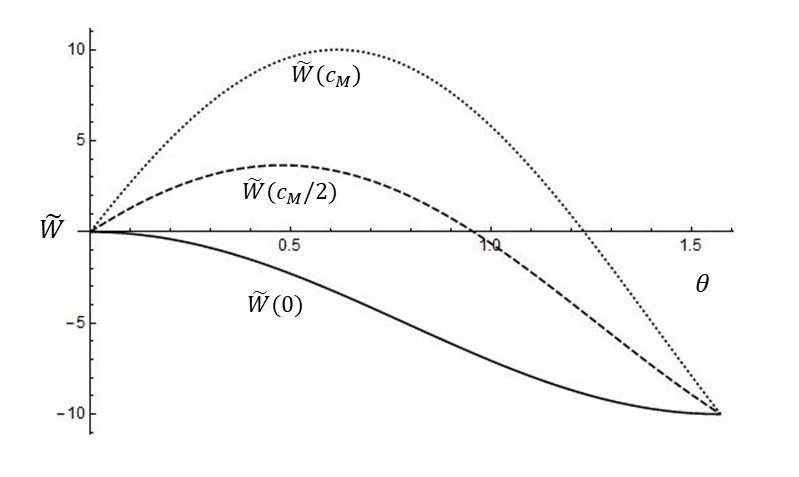}}
\caption{\label{f:fig5} $\tilde{W}$ as function of $\theta$ for
  classically correlated thermal states with no coherent signal, $N_A =
  10$, $N_B=5$, and $\omega_B=2\omega_A$, for different values of $c$,
  as in Fig. 1.  Since $N_A>N_B$, to get $\tilde{W}>0$, $ \theta $ must
  be smaller than a threshold that depends on $N_A$, $N_B$ and
  $c$ [see Eq. (\ref{theta1})]. When the modes are uncorrelated ($c=0$)
  the work extraction scheme is useless.}
\end{figure}

 When $\omega_B < \omega_A$, again the sign of
$c$ fixes $\varphi $ at $0$ or $\pi$. For $N_B < N_A$, 
we have $\tilde{W} >0$ for any $\theta$.  
Otherwise, if $N_B >
N_A$, the net increase $\tilde{W}$ is positive for
\begin{equation}
\tan \theta < \frac{2|c|}{N_B-N_A}.
\label{theta2}
\end{equation}
The optimal value of $\theta$ is provided by
\begin{equation}
\theta_{\max} = \frac{1}{2} \arctan \left( \frac{2|c|}{N_B-N_A} \right).\label{theta4}
\end{equation}
Notice that both solutions (\ref{theta3}) and (\ref{theta4}) satisfy
the respective conditions (\ref{theta1}) and (\ref{theta3}).  In
summary, for all values of $N_A$,$N_B$, and $c$, we can provide a
transformation that increases the net free energy of the system,
i.e. $\tilde{W}>0$, for suitable choice of $\varphi$ and $\theta$. \\ Finally,
note that for $c=0$ the procedure runs only for $(\omega_B
-\omega_A)(N_B -N_A) >0$, consistently with Eq. (\ref{poss}), with optimal
$\theta _{\max}=\pi/2$.  In Figs. \ref{f:fig4}
and \ref{f:fig5} we report examples of $\tilde{W}(\theta)$ as a
function of $\theta$ for different values of $c$, for fixed $N_A$, $N_B$,
and $\omega_B/\omega_A$.
\subsubsection{Exploiting type-II correlated states}
We consider the type-II correlated states, whose covariance matrix is
given in \eqref{CM classical/quantum}. We remind that the net increase
in the free energy of the system reads as $\tilde{W} =
\frac{\omega_B}{\omega_A} \Delta E_A - \Delta B_A $. For this class of
states the variation of the internal energy is obtained by Eq.  \eqref{QC
  internal energy}, namely 
\begin{equation}
\Delta E_{A} = \omega_{A}\sin^{2}\theta\left(N_{B}-N_{A}+|\delta |^2\right).
\end{equation}
The final covariance matrix of the system is given by 
\begin{eqnarray}
\sigma_{A}'= \left ( \frac{2N_{B}+1}{2}\sin^2\theta
+\frac{2N_{A}+1}{2}\cos^{2}\theta \right ) \mathbb{I}_2  +
c\sin2\theta \, {\bar R}_\varphi\,,
\end{eqnarray}
and hence the corresponding variation of the bound energy writes
\begin{eqnarray}
  \Delta B_A=
  \omega_{A}\sqrt{\left(N_{B}\sin^{2}\theta+N_{A}\cos^{2}\theta+\frac{1}{2}\right)^{2}-\left(c\sin2\theta\right)^{2}}
  - \omega_{A}\frac{2N_{A}+1}{2}\,
\end{eqnarray}
The net increase in the free energy of the system can be therefore expressed as
\begin{eqnarray}
\tilde{W} = \omega_{B}\sin^{2}\theta\left(N_{B}-N_{A}+ |\delta
|^2\right) + \omega_{A}\frac{2N_{A}+1}{2} -\omega_{A}\sqrt{\left(N_{B}
  \sin^{2}\theta+N_{A}\cos^{2}\theta+\frac{1}{2}\right)^{2}-\left(c\sin2\theta\right)^{2}}\,.
\label{wtb}
\end{eqnarray}
For fixed values of state parameters $N_A,N_B,c,\delta $
the optimal value of $\theta $ is chosen to maximize
$\tilde W$, possibly with $\tilde W _{\max}>0$.
For a better understanding of this case, let us explicitly
consider a two-mode squeezed vacuum state, which represents a type-II pure entangled state
between system and bath.  The pertaining covariance matrix is obtained
for  $\sigma_A =\sigma_B= \frac{\cosh 2r}{2}\mathbb{I}_2 $ and $c =
\frac{\sinh 2r}{2}$, along with $\delta =0$.  Then Eq. (\ref{wtb})
rewrites as
\begin{equation}
\tilde{W} = \omega_A \frac{\cosh 2r}{2} - \omega_A \sqrt{\frac{\cosh^2
    2r}{4}-\frac{\sinh^2 2r}{4}\sin^2 2\theta}.
\end{equation}
The maximal $\tilde{W}$ is attained when $ \theta = \frac{\pi}{4}$,
for which 
\begin{equation}
\tilde{W}_{\max} = \omega_A \sinh^2 r.  
\end{equation}
We notice that such an optimal transformation corresponds to a
complete removal of correlations, which are used to increase the free
energy of the system (and the bath). This effect can also be 
understood by means of the identity (\ref{alg}). In fact, since the
two-mode squeezed state is generated by $U_{PA}(\xi )$ on the vacuum
state $|0 \rangle _A \otimes |0 \rangle _B$ and $U_{FC}(\zeta ) |0
\rangle _A \otimes |0 \rangle _B =0$, one has
\begin{eqnarray}
U_{FC} \left (\frac \pi 4 \right ) U_{PA}(r)|0 \rangle _A \otimes |0
  \rangle _B =
  (S_A(r) |0 \rangle _A )\otimes (S_B (-r) |0 \rangle _B )
    \;.
\end{eqnarray}
\subsection{Parametric amplifier}
The variation of the internal energy of the bath is given by
\begin{eqnarray}
  \Delta E_{B} =  \frac {\omega_{B}}{2}
\sinh^{2}r
  \left(\Tr\left[\sigma_{A}\right]+\Tr\left[\sigma_{B}\right] + 2\lvert \delta \rvert ^2   \right)
 + \frac {\omega_{B}}{2} \sinh2r \Tr[\tilde{R}_{\psi}\epsilon ]
 \,,
\end{eqnarray}
whereas the final covariance matrix of the system reads
\begin{eqnarray}
\sigma'_{A}=&\cosh^{2}{r}
\sigma_{A}+
\sinh^{2}r\tilde{R}_{\psi}\sigma_{B}\tilde{R}_{\psi}
+& \frac{1}{2}\sinh{2r}
\left(\epsilon^{T}\tilde{R}_{\psi}+\tilde{R}_{\psi}\epsilon\right)\,.
\end{eqnarray}
For this process we have $\Delta E_B = \frac{\omega_B}{\omega_A}
\Delta E_A$ \cite{notaw,notafield}, and hence $\tilde W =
-\left(\frac{\omega_B}{\omega_A}\Delta E_A + \Delta B_A \right)$,
which can also be expressed as $\tilde W =
-\frac{\omega_B}{\omega_A}\Delta F_A - \left
(1+\frac{\omega_B}{\omega_A}\right) \Delta B_A$. In order to increase
the extractable work ($\Delta F_A>0$) we need $\Delta B_A<0$ and $|\Delta B_A | >
\frac{\omega_B}{\omega_A +\omega_B} \Delta F_A$.
\par We notice, however, that for Gaussian factorized input states one
always has $\Delta B_A >0$. In fact, for symmetry reasons $\Delta B_A
=\Delta Q$ as in Eq. (\ref{cond2}) with $\omega _A$ replacing $\omega
_B$, and hence the presence of initial
correlations is needed to obtain $\tilde W >0$. Let us consider then
the last example.
\subsubsection{Exploiting type-II correlated states}
We consider the type-II correlated states, whose covariance matrix is
given in \eqref{CM classical/quantum}.
The variation of the internal energy is obtained by
Eq. \eqref{Internal energy frequency am-1-1} with $\alpha =0$,
namely 
\begin{eqnarray}
  \Delta E_{A} = \omega_{A}\sinh^{2} r \left(N_{A}+N_{B}+1
  +|\delta
  |^2\right) 
  + \omega _A \sinh 2r \, c \,\cos \psi 
\;,
\end{eqnarray}
whereas $\Delta B _A=\Delta Q$ as in Eq. (\ref{qbb}) upon replacing
$\omega _B$ with $\omega _A$. Then one has
\begin{eqnarray}
 \tilde{W} = -
  \omega_{B}\sinh^{2}r |\delta |^{2}  - (\omega _A+\omega _B) 
  \left[\left(N_{A}+N_{B}+1\right)\sinh^{2}r+c\sinh2r\cos\psi\right]\,. 
\end{eqnarray}
For fixed values of state parameters $N_A,N_B,c,\delta $, 
the optimal values of $r$ and $\psi $ are chosen to maximize
$\tilde W$. For simplicity, let us consider the case
$\omega = \omega _A =\omega_ B$, $N_A=N_B=N$, and $c=\sqrt{N(N+1)}$, which
corresponds to an input pure two-mode squeezed state displaced by a
coherent signal $\delta $ in the bath mode. Then,
\begin{eqnarray}
\tilde W = -\omega  [(4N+2+|\delta |^2)\sinh ^2 r
  +2\sqrt{N(N+1)}\sinh 2r \cos \psi ]
  \;.
\end{eqnarray}
Clearly, the optimal choice for $\psi $ is $\psi =\pi$, and by solving
$\partial \tilde W /\partial r =0$ one obtains the optimal value of
$r$ as $r_{\max}=\mbox{atanh} \frac{2\sqrt {N(N+1)}}{4N+2+|\delta |^2}$. The
corresponding optimal net increase of the system free energy is given
by
\begin{eqnarray}
  \tilde W _{\max}= \omega \frac{4N (N+1)(4N+2+|\delta  |^2)}
         {(4N+2+|\delta  |^2)^2-4N(N+1)}\;.
\end{eqnarray}
\section{Conclusions} 
In this work we have studied the thermodynamics of two bosonic systems
that interact via entropy-preserving transformations in the mode
operators. The first mode represents the thermodynamical system, while
the second describes the bath which, differently from standard
formulations of thermodynamics, is treated as a quantum system, namely
it can possess coherence or squeezing, and can be correlated with the
system. The main result of this work is the formulation of the first
law of thermodynamics for any two-mode states, hence the balance
between the heat (\ref{Heat bosonic}) and the variation of the
internal energy \eqref{Internal Energy2} of the system which gives the
work performed onto the thermodynamical system in the entropy
preserving transformation.  We have systematically considered the two
different bilinear transformations, namely the frequency
converter/beam splitter and the parametric amplification. Although our
results hold for any two-mode states, we have mainly focused on the
case of Gaussian states, where the heat and the variation of the
internal energy take simple expressions. In particular, we have
derived the first law in the case of initial uncorrelated modes,
providing the most general formulation for bilinear
transformations. Moreover, we have considered the case of initial
correlations between the thermodynamical system and the bath by
analyzing two types of correlated states, one that considers only
separable states and the other that enables also entanglement.  The
case of correlated states has also been considered from an
information-theoretical point of view by means of the Renyi entropy of
order two, thus showing how anomalous heat flows can occur by
exploiting correlations. Finally, we have proposed work-extraction
schemes, showing how one can engineer entropy-preserving
transformations to increase the free energy of the system, namely the
amount of extractable work, by letting the system interact with the
bath, thus exploiting the presence of correlations, squeezing or
coherence.


\begin{thebibliography}{99}
\bibitem{foot} The determinant of $A+B$, where A and B are $2 \times
  2$ symmetric matrices, can be expressed as: $\det\left( A+B
  \right)= \det A +\det B +\widetilde{AB}$, with $ \widetilde{AB}
  =A_{11}B_{22}+A_{22}B_{11}-2A_{12}B_{12}$, where $X_{ij}$ denotes
  the matrix element of $X$.
\bibitem{notaw} In frequency conversion 
$a^\dag a + b^\dag b$ is a constant of motion, as long as the
  parametric approximation holds \cite{parap}. Hence, $W=(\omega _A
  -\omega _B) \langle \Delta a^\dag a \rangle $ represents the energy
  absorbed (for $W >0$) or released (for $W<0$) to the classical pump
  at frequency $\omega _P =|\omega _A - \omega _B|$. Analogously, in parametric
  amplification $a^\dag a - b^\dag b$ is a constant of motion, and
  $W=(\omega _A +\omega _B) \langle \Delta a^\dag a \rangle $, with pump frequency
  $\omega _P =\omega _A + \omega _B$.
\bibitem{notafield}Remarkably, in the case of parametric amplification
  the role of the classical pump as a work reservoir is analogous to that of the
  driving classical spacetime as the source for the change in the internal
  energy of quantum fields developed in Ref. \cite{Fields1}.
\end{thebibliography}
\end{document}